%% file: pap4000.tex
\documentclass{aa}
\usepackage{txfonts}
\usepackage{graphicx}
\begin{document}
   \title{Using spectroscopic data to disentangle\\
          stellar population properties}


   \author{N.~Cardiel \inst{1,2}
           \and
           J.~Gorgas \inst{2}
           \and
           P.~S\'{a}nchez-Bl\'{a}zquez \inst{2}
           \and \\
           A.~J. Cenarro \inst{2}
           \and
           S.~Pedraz \inst{1,2}
           \and
           G.~Bruzual \inst{3}
           \and
           J.~Klement \inst{4}
          }

   \offprints{N. Cardiel,\\
              \email{ncl@astrax.fis.ucm.es}}

   \institute{Calar Alto Observatory, CAHA, Apdo.~511, 04004, Almer\'{\i}a,
              Spain
         \and
              Departamento de Astrof\'{\i}sica, Facultad de F\'{\i}sicas,
              Universidad Complutense de Madrid, E28040-Madrid, Spain
         \and
              Centro de Investigaciones de Astronom\'{\i}a (CIDA), Apartado
              Postal 264, M\'{e}rida 5101-A, Venezuela
         \and
              Institut f\"{u}r Astronomie, ETH Zentrum, SEC E3,
              Scheuchzerstrasse 7,  CH-8092 Z\"{u}rich
             }

   \date{Received February ??, 2003; accepted June ??, 2003}

   \abstract{It is well known that, when analyzed at the light of current
   synthesis model predictions, variations in the physical properties of single
   stellar populations (e.g.\ age, metallicity, initial mass function, element
   abundance ratios) may have a similar effect in their integrated spectral
   energy distributions. The confusion is even worsened when more realistic
   scenarios, i.e.\ composite star formation histories, are considered. This
   is, in fact, one of the major problems when facing the study of stellar
   populations in star clusters and galaxies.  Typically, the observational
   efforts have been aimed to find the most appropriate spectroscopic
   indicators in order to avoid, as far as possible, degeneracies in the
   parameter space. However, from a practical point of view, the most suited
   observables are not, necessarily, those that provide more orthogonality in
   that parameter space, but those that give the best balance between parameter
   degeneracy and sensitivity to signal-to-noise ratio per \AA, ${\mbox{\rm
   {\it SN}({\rm \AA})}}$. In order to achieve the minimum combined total error
   in the derived physical parameters, this work discusses how the
   functional dependence of typical line-strength indices and colors on
   ${\mbox{\rm {\it SN}({\rm \AA})}}$ allows to define a {\it suitability
   parameter\/} which helps to obtain better realistic combinations of
   spectroscopic data.  As an example, we discuss in more detail the problem of
   breaking the well known age-metallicity degeneracy in relatively old stellar
   populations, comparing the suitability of different spectroscopic 
   diagrams 
   for a simple stellar population of solar metallicity and 12~Gyr old.
    }

   \maketitle
%
\section{Introduction}

\subsection{Evolutionary synthesis modeling}
\label{subsection_intro_models}

In order to predict the expected spectral energy distribution (SED) of simple
stellar populations (chemically homogeneous and coeval stellar systems), it is
possible to use first principles (e.g.\ initial mass function, star formation
rate, stellar isochrones, element abundance ratios) to generate synthetic star
systems.  This technique, known as evolutionary synthesis modeling, has been
widely employed to understand the origin and evolution of star clusters and
galaxies (Crampin \& Hoyle \cite{crampin61}; Tinsley \cite{tinsley72},
\cite{tinsley78}, \cite{tinsley80}; Tinsley \& Gunn \cite{tinsley76}; Gunn,
Stryker \& Tinsley \cite{gunn81}; Bruzual \cite{bruzual83}, \cite{bruzual02};
Arag\'{o}n-Salamanca, Gorgas \& Rego \cite{aragon87}; Charlot \& Bruzual
\cite{charlot91}; Bruzual \& Charlot \cite{bruzual93}; Arimoto \& Yoshii
\cite{arimoto86}, \cite{arimoto87}; Guiderdoni \& Rocca-Volmerange
\cite{guiderdoni87}; Buzzoni \cite{buzzoni89}, \cite{buzzoni95}; Mas-Hesse \&
Kunth \cite{mmas91}; \mbox{Fritze-v}.\ Alvensleben \& Gerhard \cite{fritze94};
Cervi\~{n}o \& Mas-Hesse \cite{cervino94}; Worthey \cite{worthey94}; Worthey \&
Ottaviani \cite{worthey97}; Bressan, Chiosi \& Fagotto \cite{bressan94};
Chiosi, Bressan \& Fagotto \cite{chiosi96}; Tantalo et al.~\cite{tantalo98};
Milone, Barbuy \& Bica \cite{milone95}; Leitherer \& Heckman
\cite{leitherer95}; Leitherer et al.~\cite{leitherer96}, \cite{leitherer99};
Fioc \& Rocca-Volmerange \cite{fioc97}; Vazdekis et al.~\cite{vazdekis96},
\cite{vazdekis97}, \cite{vazdekis03}; Vazdekis \cite{vazdekis99}; Mayya
\cite{mayya95}, \cite{mayya97}; Garc\'{\i}a-Vargas, Moll\'{a} \& Bressan
\cite{marisa98}; Moll\'{a} \& Garc\'{\i}a-Vargas \cite{molla00}; Maraston
\cite{maraston98}; Schiavon, Barbuy \& Bruzual \cite{schiavon00}; Origlia \&
Oliva \cite{origlia00}; Zackrisson et al.~\cite{zackrisson01}; Thomas, Maraston
\& Bender \cite{thomas03}).

The reliability of model predictions has greatly increased as their developers
include more realistic physical ingredients. However, as discussed by Charlot,
Worthey \& Bressan (\cite{charlot96}), there are still problems due to
uncertainties in the theory of stellar evolution (e.g.\ post-main-sequence
stages), the physics of stellar interiors (e.g.\ atomic diffusion, helium
content, the temperature of the red giant branch), and the lack of complete
stellar spectra libraries.  It is important to note that although initially it
is straightforward to predict spectroscopic indices from this type of models,
the inherent problems associated to the SED libraries, either empirical or
theoretical, have a non negligible influence in the line-strength predictions.
For instance, empirical SED libraries constitute a coarse grained, and usually
incomplete (specially for nonsolar metallicities and nonsolar abundance
ratios) sampling of the atmospheric stellar parameter space, whereas
theoretical libraries usually exhibit systematic discrepancies among themselves
and when compared with observational data (e.g.\ Lejeune, Cuisinier \& Buser
1997, 1998). 

The use of empirical fitting functions (e.g.\ Gorgas et al.~\cite{gorgas93},
\cite{gorgas99}; Worthey et al.~\cite{wortheyetal94}; 
Cenarro et al.~\cite{cenarro02}) 
can help to reduce substantially
these effects (Worthey \cite{worthey94}; Vazdekis et al.~\cite{vazdekis03}). 
They do not only allow the computation of line-strength indices
for any given combination of input parameters, but the error in their
predictions can be minimized with the use of a large set of stars.  However,
and since the empirical fitting functions only predict the value of a given
line-strength feature for a fixed set of stellar atmospheric parameters, the
shape of the spectrum that leads to such value is therefore unknown. To insert
the fitting function predictions into the evolutionary synthesis models it is
necessary to use the local continuum of each single star in the SED library as
a reference continuum level. In this way it is possible to weight the
luminosity contribution of each type of star, in the neighborhood wavelength
region of each index, to obtain the final line-strength prediction.

In addition, there are also additional sources of biases in model predictions.
Cervi\~{n}o et al.~(\cite{cervino00}, \cite{cervino01}, \cite{cervino02}), and
Cervi\~{n}o \& Valls-Gabaud~(\cite{cervino03})
have thoroughly analyzed the impact of the actual
discreteness of real stellar populations (see also Bruzual \cite{bruzual01},
and references therein), the Poissonian dispersion due to finite populations in
non-time-integrated observables, and the influence of the interpolations in
time-integrated quantities, among others.

But far from being a discouraging situation, the recognition of all these
problems is providing a solid understanding of the challenging task of modeling
stellar populations. In this sense, the collective effort of many modelers
(e.g.  Leitherer et al.~\cite{leitherer96}) is given strength to the idea that
reliable and unbiased model predictions are starting to emerge.

\subsection{Physical parameter degeneracy}

Although spectroscopic data provide a direct way to analyze the integrated
light of composite stellar systems, the predictions from simple stellar
population synthesis models reveal that variations in the relevant physical
properties of such systems may produce quite similar spectral energy
distributions (SEDs).  This {\it conspiracy\/} leads to undesirable
degeneracies when passing from the observable space (e.g. that defined by
line-strength indices and colors), to the parameter space (age, metallicity,
initial mass function, etc.).

Among the best known examples of degeneracy we must highlight the one exhibited
by age and metallicity in the study of relatively old stellar populations
(O'Connell \cite{oconnell76}, \cite{oconnell80}, \cite{oconnell94}; Aaronson et
al. \cite{aaronson78}; Worthey \cite{worthey94}; Faber et al.~\cite{faber94}).
This outstanding problem drove many authors to seek for spectral line-strength
indices which were more sensitive to age than to metallicity and vice versa
(e.g. Rose~\cite{rose85}, \cite{rose94}; Worthey~\cite{worthey94}). In this
sense, Worthey (1994) introduced an interesting quantitative measure of the
metal sensitivity of each index, computed as the partial derivatives
$d\log({\rm age})/d\log({\rm Z})$ around his model predictions for a 12~Gyr old
stellar population with solar metallicity. Since then, large efforts have been
focused toward the search of spectral features with very high (e.g.\ Fe4668)
and very low (e.g.\ H$\beta$, H$\gamma$) metal sensitivities. However, this
work has led to the use of individual and narrow absorption features (e.g.\
Jones \& Worthey \cite{jones95}; Worthey \& Ottaviani \cite{worthey97};
Vazdekis \& Arimoto \cite{vazdekis99}) for which accurate measurements demand
high signal-to-noise ratios. In addition, these spectral signatures are usually
very sensitive to spectral resolution and, therefore, velocity dispersion.

\subsection{Compromise between orthogonality and errors}

It is important to note that since the problem is to break a degeneracy, in
practice the real concern is how uncertain the requested physical parameters
are when derived from a particular observable space.  In this sense, two
circumstances have to be carefully handled. The first is the orthogonality of
the iso-parameter lines in the observable space.  As we have just mentioned,
this is precisely the major concern of previous works. The second condition to
be aware of is the propagation of the errors in the spectroscopic indices into
the corresponding errors in the parameters.  However, and as it is expected,
narrow indices (better suited to provide more orthogonality) exhibit larger
errors than broad spectral features, for a given signal-to-noise ratio.
Summarizing, orthogonality and small errors are magnitudes that can not be, a
priori, simultaneously maximized. As a result, it seems clear that the most
suited observable space will be that in which the two mentioned requirements
are best balanced.

The relevance of finding this equilibrium can hardly be overemphasized,
specially when one considers the important observational effort that is being
(or is going to be) spent in ambitious spectroscopic surveys, like e.g.\ DEEP
(Mould~\cite{mould93}; Koo~\cite{koo95}), EFAR (Wegner et al.~\cite{wegner96}),
CFRS (Lilly et al.~\cite{lilly95}; Hammer et al.~\cite{hammer97}), Sloan (York
et al.~\cite{york00}; Kauffmann et al.~\cite{kauffmann03}), VLT-VIRMOS (Le
F\`{e}vre et al.~\cite{lefevre00}), SDSS (Bernardi et
al.~\cite{bernardi03}). In all these type of
surveys, a large amount of spectroscopic data is collected, although
signal-to-noise ratios and spectral resolution are typically moderate. These
factors strengthen the need of a quantitative estimation of the reliability of
the physical parameters derived from such spectroscopic studies.

With a clear practical sense, in this paper we explore the way to determine
those combinations of spectroscopic observables that provide robust tools to
constrain physical properties in stellar populations. For this purpose, we
are going to assume that evolutionary synthesis model predictions are error
free. Although, as we have discussed in Sect.~\ref{subsection_intro_models},
this is not at present the case, we want to concentrate in the problem of
balancing errors and degeneracy. For this reason, model uncertainties and
biases are out of the scope of this paper. In
section~\ref{section_errors_in_spectroscopic_measurements} we review and
enlarge our previous work concerning random errors in line-strength indices,
showing that a common functional dependence of final index errors on
signal-to-noise ratio can be found for different index and color definitions.
In section~\ref{section_reliability_of_physical_properties} we obtain simple
formulae to quantify total errors in the derived physical parameters when
derived from spectroscopic measurements. As an illustrative example, we examine
in more detail the age-metallicity degeneracy in
section~\ref{section_age_metallicty_degeneracy}, through the comparison of
the suitability of different spectroscopic diagrams for a 12~Gyr old simple
stellar population with solar metallicity. We summarize the conclusions of
this work in section~\ref{section_conclusions}. Finally, some more technical
aspects have been deferred to
Appendices~\ref{appendix_d4000}--\ref{appendix_sensitivity}.

%
\section{Errors and spectroscopic measurements}
\label{section_errors_in_spectroscopic_measurements}

\subsection{Error handling}
\label{sect_error_handling}

Random uncertainties and biases are inherently associated to the physical
process of data acquisition. Random errors can be easily derived with the help
of statistical methods. Unfortunately, the situation is not so simple when
handling systematic effects, where a case by case solution must be sought. In
practice, the aim is to obtain reliable quantitative constraints of the total
random errors present in the data while having uncorrected systematic effects
(if any) well within the range spanned by the former. For this to be the case,
possible sources of systematic effects should be identified and alleviated
during the measure process. In this paper we are assuming that this is actually
the case, and for that reason we are exclusively focusing on the impact of
random errors.

Although, as we have just mentioned, appropriate observational strategies can
greatly help in reducing the sources of data biases, the unavoidable limited
exposure time that can be spent in each target determines the maximum
signal-to-noise ratio in practice achievable. The data reduction process, aimed
to minimize the impact of data acquisition imperfections on the measurement of
data properties with a scientific meaning for the astronomer, is typically
performed by means of arithmetical manipulations of data and calibration
frames. As a result, the initial random errors present in the raw scientific
and calibration data are combined (and thus enlarged) and propagated
throughout the reduction procedure. 

In a recent paper, Cardiel et al.~(\cite{cardiel02}) have discussed the
benefits and drawbacks of different methods to quantify random errors in the
context of data reduction pipelines. One of the conclusions of this work is
that a parallel reduction of data and error frames is likely the most elegant
and general approach, and, in some circumstances, the only way to proceed when
observing or computing time demands are specially restrictive. It must be
noted, however, that in order to apply this method to compute final errors, it
is essential to prevent the introduction of correlation between neighboring
pixels, which dangerously leads to underestimated errors. This problem arises
when one performs image manipulations involving rebinning or non-integer pixel
shifts of data, which is the case of those data reduction steps devoted to
correct for geometric distortions, to produce a wavelength calibration into a
linear scale, or to correct for differential refraction variations with
wavelength, to mention a few. Fortunately, a modification in the way typical
reduction processes operates can help to solve this problem. Although we are
not going to enter into details (we refer the interested reader to that paper),
the key point is to transfer the responsibility of the most complex reduction
steps to the analysis tools, which must manipulate data and error frames using
a distorted system of coordinates, overriding the orthogonal coordinate system
defined by the physical pixels in a detector.

\subsection{Random errors in spectroscopic measurements}

Once it can be assumed that reliable final random error estimates are
available, and that, in comparison, systematic biases are not relevant, it is
straightforward to obtain a quantitative estimate of the error in a given
spectroscopic measurement. Since the information collected by detectors is
physically sampled in pixels, the starting point in the analysis of a single
spectrum will be the spectrum itself $S(\lambda_i)$ (with 
$i=1,...,N_{\rm pixels}$) and
its associated random error spectrum $\sigma_S(\lambda_i)$. In the following
discussion, we are assuming
\begin{eqnarray}
{\rm cov}[S(\lambda_i),S(\lambda_j)] & = & 
  \langle S(\lambda_i)\;S(\lambda_j) \rangle - 
  \langle S(\lambda_i) \rangle \; \langle S(\lambda_j) \rangle = 0, \nonumber\\
 &   & \mbox{}\hfill \forall i,j \in [1,N_{\rm pixels}], \;\; i \neq j,
\end{eqnarray}
i.e.\ errors are not correlated. From here, two different approaches can be
followed (Cardiel et al.~\cite{cardiel98}; Cardiel~\cite{cardiel99}). 

One possibility is to estimate numerically the effect of errors via Monte Carlo
simulations. In practice, new instances of the spectrum,
$\tilde{S}(\lambda_i)$, can be generated introducing Gaussian noise in each
pixel using, for example,
\begin{equation} 
\tilde{S}(\lambda_i) = S(\lambda_i) 
            + \sqrt{2} \; \sigma_S(\lambda_i) \; 
              \sqrt{-\ln(1-r_1)} \; \cos(2 \pi r_2),
\label{equation_randomspectrum}
\end{equation}
where $r_1$ and $r_2$ are two random numbers in the range $r_1,r_2 \in [0,1)$.
The unbiased standard deviation of all the measurements, performed over a
sufficiently large number of simulated spectra, provides the final error.

Another method to estimate errors consists in the use of analytical formulae.
In fact, Cardiel et al.~(\cite{cardiel98}) and Cenarro et
al.~(\cite{cenarro01}) have already presented analytical expressions to compute
errors in the 4000~\AA\ break (D$_{4000}$; defined by
Bruzual~\cite{bruzual83}), and in atomic, molecular and generic indices.
Interestingly, in the case of molecular indices, and when atomic and generic
indices are measured in magnitudes, index errors can be derived by (see e.g.
Appendix~A in Cenarro et al.~\cite{cenarro01})
\begin{equation}
\sigma(M) \simeq \frac{c(M)}{\mbox{\rm {\it SN}({\rm \AA})}},
\label{equation_error_sn}
\end{equation}
where $\sigma(M)$ is the random error in the index $M$, $\mbox{\rm {\it
SN}({\rm \AA})}$ is the averaged signal-to-noise ratio per \AA\ measured in the
pixels included in all the bandpasses which define the considered index, and
$c(M)$ is an index dependent constant.  A similar expression can be deduced for
the D$_{4000}$ (Cardiel~\cite{cardiel99}), namely
\begin{equation}
\sigma({\rm D}_{4000}) \simeq
   \frac{{\rm D}_{4000}}{10}
   \frac{1}{\mbox{\rm {\it SN}({\rm \AA})}}.
\label{equation_error_sn_old_d4000}
\end{equation}
However, in order to take advantage of the fact that the relative error in
D$_{4000}$ is exclusively a function of $\mbox{\rm {\it SN}({\rm \AA})}$, it 
is more useful to measure this kind of spectral feature in
logarithmic units, i.e.
\begin{equation}
\tilde{\rm D}_{4000} \equiv 2.5\;\log_{10}({\rm D}_{4000}),
\end{equation}
since, with this definition,
\begin{equation}
\sigma(\tilde{\rm D}_{4000}) \simeq
   \frac{2.5\;\log_{10}{\rm e}}{10}\;
   \frac{1}{\mbox{\rm {\it SN}({\rm \AA})}} \equiv
   \frac{c(\tilde{\rm D}_{4000})}{\mbox{\rm {\it SN}({\rm \AA})}},
\label{equation_error_sn_d4000}
\end{equation}
which has the same functional form that Eq.~(\ref{equation_error_sn}). However,
since the definition of this index covers a relatively large wavelength range,
the computation of the averaged signal-to-noise ratio in the considered
bandpasses may introduce a systematic bias in the estimation of the random
error. This point is treated in more detail in Appendix~\ref{appendix_d4000}.
In addition, in Appendix~\ref{appendix_colors} we show that
Eq.~(\ref{equation_error_sn}) also holds when using colors.

In summary, errors in typical spectroscopic measurements (line-strength indices
and colors) can be accurately estimated as a constant divided by an appropriate
average of the signal-to-noise ratio per \AA. Table~\ref{table_c_coefficients}
lists those constants for common line-strength features and colors.
Note that for colors we have employed Eq.~(\ref{equation_xi1xi2}). For
classical molecular indices (and atomic indices measured in magnitudes), which
are defined with the help of three bandpasses, the error coefficients are
computed as (see Eqs.~(44) and~(45) in Cardiel et al.~\cite{cardiel98})
\begin{equation}
\begin{array}{l}
c(M) = 2.5 \; \log_{10}({\rm e}) \times \\ \noalign{\medskip}
\;\;\; \sqrt{
       \displaystyle\frac{1}{\Delta\lambda_c}+
       \left( \frac{\lambda_r-\lambda_c}{\lambda_r-\lambda_b} \right)^2
       \displaystyle\frac{1}{\Delta\lambda_b}+
       \left( \frac{\lambda_c-\lambda_b}{\lambda_r-\lambda_b} \right)^2
       \displaystyle\frac{1}{\Delta\lambda_r}
       },
\end{array}
\label{coefficients_vs_bandpass_width}
\end{equation}
where $\lambda_b$, $\lambda_c$ and $\lambda_r$ are the central wavelengths of
the blue, central and red bandpasses, respectively, and $\Delta\lambda_b$,
$\Delta\lambda_c$ and $\Delta\lambda_r$ are the bandpass widths\footnote{For
those readers interested in computing expected errors in atomic indices
measured in \AA, it is possible to use
\begin{displaymath}
\sigma(I_{\rm a}) \simeq 
  \frac{c_1 - c_2 \; I_{\rm a}}{{\mbox{\rm {\it SN}({\rm \AA})}}},
\end{displaymath}
where $c_1$ and $c_2$ coefficients for typical line-strengh features are given
in Table~1 of Cardiel et al.~(\cite{cardiel98}). Note that in this case the
error estimates depend on the absolute index value.}. 

\begin{figure}
 \resizebox{1.0\hsize}{!}{\includegraphics{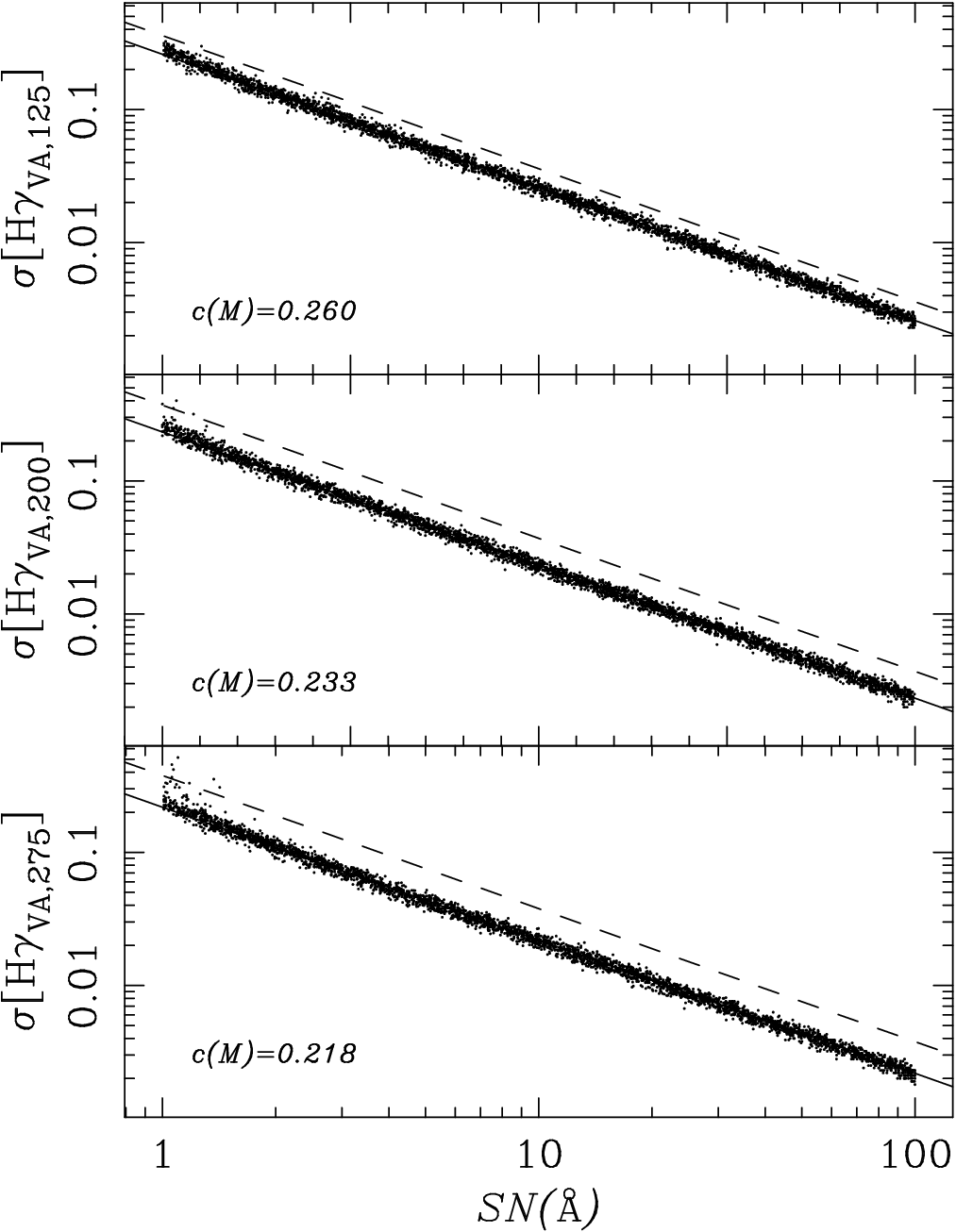}}
 \caption{Random errors from numerical simulations in the measurement of
 the H$\gamma_{{\rm VA},\sigma}$ indices defined by Vazdekis \& Arimoto
 (\cite{vazdekis99}), as a function of the mean signal-to-noise ratio per \AA. 
 The fit to the data points (full line) provides the error
 coefficients $c(M)$ defined in Eq.~(\ref{equation_error_sn}). The dashed line
 indicates what should be the location of the data if the bandpasses defining
 those indices did not overlap.}
 \label{plot_c3HgVA}
\end{figure}

It is important to highlight that the coefficients $c(M)$ obtained with the
previous expression are valid as long as the three bandpasses do not overlap.
If this is not the case, the coefficients can be computed numerically through
numerical simulations. This is in fact the situation for the three narrow
$H\gamma_{{\rm VA},\sigma}$ indices introduced by Vazdekis \& Arimoto
(\cite{vazdekis99}), in which part of the central bandpass containing the
spectral feature overlaps with the blue bandpass of the pseudo-continuum. In
Fig.~\ref{plot_c3HgVA} we show the result of simulating 4000 spectra using
Eq.~(\ref{equation_randomspectrum}), from which we have derived the $c(M)$
coefficients for H$\gamma_{{\rm VA},\sigma}$ by fitting the measured random
errors as a function of $\mbox{\rm {\it SN}({\rm \AA})}$.

\begin{table}[t]
\centering
\begin{tabular}{lc|lc|lc} \hline
index & $c$ & index & $c$ & color & $c$ \\ \hline\noalign{\smallskip}
$\tilde{\rm D}_{4000}$                  &  0.109 &
Fe4531                                  &  0.273 &  
$(u-g)_{\mbox{\scriptsize AB}}$         &  0.0500\\ 
H$\delta_{\mbox{A}}$                    &  0.217 &
Fe4668                                  &  0.224 &
$(g-r)_{\mbox{\scriptsize AB}}$         &  0.0421\\
H$\delta_{\mbox{F}}$                    &  0.281 &
H$\beta$                                &  0.276 &
$(g-i)_{\mbox{\scriptsize AB}}$         &  0.0412\\
CN$_1$                                  &  0.224 &
Fe5015                                  &  0.234 &
$(g-z)_{\mbox{\scriptsize AB}}$         &  0.0393\\
CN$_2$                                  &  0.269 &
Mg$_1$                                  &  0.166 &
$(g-J)_{\mbox{\scriptsize AB}}$         &  0.0368\\
Ca4227                                  &  0.400 &
Mg$_2$                                  &  0.193 &
$(g-H)_{\mbox{\scriptsize AB}}$         &  0.0354\\
G4300                                   &  0.265 &
Mgb                                     &  0.268 &
$(g-Ks)_{\mbox{\scriptsize AB}}$        &  0.0348\\
H$\gamma_{\mbox{A}}$                    &  0.204 &
Fe5270                                  &  0.251 &
$(U-B)$                                 &  0.0502\\
H$\gamma_{\mbox{F}}$                    &  0.274 &
Fe5335                                  &  0.292 &
$(B-V)$                                 &  0.0434\\
H$\gamma_{\mbox{HR}}$                   &  0.887 &
Fe5406                                  &  0.314 &
$(V-R)$                                 &  0.0373\\
H$\gamma_{\mbox{\scriptsize VA,125}}$   &  0.260 &
Fe5709                                  &  0.291 &
$(V-I)$                                 &  0.0367\\
H$\gamma_{\mbox{\scriptsize VA,200}}$   &  0.233 &
Fe5782                                  &  0.333 &
$(V-J)$                                 &  0.0391\\
H$\gamma_{\mbox{\scriptsize VA,275}}$   &  0.218 &
Na5895                                  &  0.271 &
$(V-K)$                                 &  0.0352\\
Fe4383                                  &  0.280 &
TiO$_1$                                 &  0.182 &
$(R-I)$                                 &  0.0285\\
Ca4455                                  &  0.340 &
TiO$_2$                                 &  0.157 &
$(J-H)$                                 &  0.0309\\
                                        &        &
                                        &        &
$(H-K)$                                 &  0.0257\\ \hline
\end{tabular}
\caption{Coefficients to estimate the expected random error in typical
line-strength indices and colors, as a function of the mean signal-to-noise
ratio per \AA, following Eq.~(\ref{equation_error_sn}). Note that these
coefficients are valid for atomic indices when measured in magnitudes (i.e.\ as
if they were molecular indices). For line-strength indices measured in
\AA\ see coefficients in Table 1 in Cardiel et al.~(\cite{cardiel98}).} 
\label{table_c_coefficients}
\end{table}

\section{Reliability of physical properties derived from spectroscopic data}
\label{section_reliability_of_physical_properties}

Given an $n$-dimensional observational space, built with the help of
$n$~spectroscopic measurements $M_i$, with
$i=1,...,n$, small variations in all those measurements around a given point
$(m_1^0,m_2^0,...,m_i^0,...,m_n^0)$ in that space can be expressed, using the
prediction of evolutionary synthesis models, as a linear function of $n$
variations of physical parameters $P_j$ around the point
$(p_1^0,p_2^0,...,p_j^0,...,p_n^0)$ of the form
\begin{equation}
\Delta m_i \simeq \sum_{j=1}^{n} a_{ij} \; \Delta p_j,
\end{equation}
being $(p_1^0,p_2^0,...,p_j^0,...,p_n^0)$ the physical parameters
associated to the observables $(m_1^0,m_2^0,...,m_i^0,...,m_n^0)$, and where
in general the $a_{ij}$ coefficients are functions of 
$(m_1^0,m_2^0,...,m_i^0,...,m_n^0)$. If the $n \times n$ matrix $A$ of the
coefficients $a_{ij}$ is not singular (i.e.\ $\det(A) \ne 0$),
we can also express locally variations in the 
physical parameters as a function of variations in the spectroscopic 
measurements by
\begin{equation}
\Delta p_j \simeq \sum_{i=1}^{n} b_{ji} \; \Delta m_i.
\label{equation_delta_sigma}
\end{equation}
Since each set of spectroscopic measurements will be, in practice, affected by
random errors $\sigma(m_i)$ (assumed to follow a normal distribution), a random
sampling of a given astronomical object provides a cloud of points of
coordinates $(m_1,m_2,...,m_i,...,m_n)_{k}$ (with $k=1,...,N_{\mbox{\scriptsize
points}}$) scattered around $(m_1^0,m_2^0,...,m_i^0,...,m_n^0)$ and following a
multivariate normal distribution.  Assuming that the bandpasses defining $m_i$
and $m_j$ do not overlap $\forall i,j$ with $i \ne j$ (i.e.\ their random
errors are uncorrelated), the surfaces of constant probability density are
hyperellipsoids centered at $(m_1^0,m_2^0,...,m_i^0,...,m_n^0)$, with axes
parallel to the coordinate axes.  The volume $V_M$ of the hyperellipsoids (for
a fixed confidence level $1-\alpha$) in the $n$-dimensional space defined by
the spectroscopic measurements will be proportional to the product of the
semiaxes (Kendall~\cite{kendall61})
\begin{equation}
V_{M} = \frac{2\;[\pi \; \chi^2_n(\alpha)]^{n/2}}{n \; \Gamma(n/2)}
  \displaystyle\prod_{i=1}^{n} \sigma(m_i),
\end{equation}
where $\chi^2_n(\alpha)$ is the upper $(100\alpha)^{\rm th}$ percentile
of a chi-square distribution with $n$ degrees of freedom, and $\Gamma(n/2)$ is
the complete gamma function.
When considering the $n$-dimensional space defined by the physical parameters
$P_j$, the above hyperellipsoids are transformed into new objects, which in
general will be no longer hyperellipsoids. However, the volume contained within
these $n$-dimensional objects can still be easily computed as
\begin{equation}
V_{P} = | \det(B) | \; V_{M},
\end{equation}
where $B$ is the $n \times n$ matrix of the coefficients $b_{ji}$.  In fact,
replacing $\Delta p_j$ and $\Delta m_i$ in Eq.~(\ref{equation_delta_sigma}) by
the errors $\sigma(p_j)$ and $\sigma(m_i)$ respectively, and using the result
given in Eq.~(\ref{equation_error_sn}), the volume $V_P$ can be obtained as
\begin{eqnarray}
V_{P} & = & | \det(B) | \; 
            \frac{2\;[\pi \; \chi^2_n(\alpha)]^{n/2}}{n \; \Gamma(n/2)}
            \displaystyle\prod_{i=1}^{n} 
              \frac{c(m_i)}{{\it SN}(\mbox{\AA})_i} =
  \nonumber \\ \noalign{\medskip}
  & = & \kappa \; \phi(\alpha,n) \;
  \displaystyle\prod_{i=1}^{n} \frac{1}{{\it SN}(\mbox{\AA})_i},
\end{eqnarray}
where
\begin{equation}
\kappa \equiv |\det(B)| \; \displaystyle\prod_{i=1}^{n} c(m_i) =
              \frac{1}{|\det(A)|} \; \displaystyle\prod_{i=1}^{n} c(m_i),
\end{equation}
and
\begin{equation}
\phi(\alpha,n) \equiv 
  \frac{2\;[\pi \; \chi^2_n(\alpha)]^{n/2}}{n \; \Gamma(n/2)}.
\end{equation}
Summarizing, for a fixed number of spectroscopic measurements, the
$\phi(\alpha,n)$ factor can be considered as a constant, and
the volume
interior to the surfaces of constant probability in the $n$-dimensional space
constructed by the physical parameters $P_j$ ($j=1,...,n$), can be expressed as
\begin{equation}
V_{P} \propto \kappa \;
              \displaystyle\prod_{i=1}^{n} \frac{1}{{\it SN}(\mbox{\AA})_i},
\end{equation}
where $\kappa$ encapsulates both the sensitivity of the spectroscopic indices
to the signal-to-noise ratio ---through the $c(m_i)$ coefficients---, and the
degree of degeneracy between the derived physical parameters ---by way of the
$a_{ij}$ or $b_{ji}$ coefficients---. For that reason $\kappa$ is a parameter
that provides quantitative information concerning the suitability of a given
set of spectroscopic indices to the study of the corresponding physical
parameters.  For our purposes, the best observational space will be that for
which this {\it suitability\/} parameter attains the lowest value (i.e.\
minimum $V_{P}$ volume). In this sense, an almost orthogonal parameter space
greatly helps to reduce this number, although its precise value will also be
strongly constrained by the signal-to-noise dependence of the measured
spectroscopic features.

%
\section{Age-Metallicity degeneracy: a case study}
\label{section_age_metallicty_degeneracy}

As an illustrative example, we can analyze the typical problem of breaking the
age-metallicity degeneracy from two-dimensional diagrams built with
line-strength indices and colors. 

\subsection{Computing the suitability parameter}

In this case, the observational space is defined by two spectroscopic
measurements $M_1$ and $M_2$. Small variations in both indices around a given
point ($m_{1}^{0},m_{2}^{0}$) can be expressed as a linear function of age and
metallicity of the form

\begin{equation}
\left(\begin{array}{@{}c@{}} \Delta m_1 \\ \Delta m_2 \end{array}\right) \simeq
\left(\begin{array}{@{}c@{\;\;}c@{}} a_{11} & a_{12} \\
                                     a_{21} & a_{22} \\ \end{array}\right)
\left(\begin{array}{@{}c@{}} \Delta\log(Z) \\ 
                             \Delta\log({\rm age}) \end{array}\right),
\label{eq_linear_approx}
\end{equation}
where, in general, the $a_{ij}$ coefficients are functions of $m_{1}^{0}$
and $m_{2}^{0}$. It is straightforward to show that
\begin{equation}
\begin{array}{@{}ll}
a_{11} \simeq S_{M_1} \; a_{12}, & {\rm and}
\\ \noalign{\medskip}
a_{22} \simeq \displaystyle\frac{a_{21}}{S_{M_2}}, & 
\end{array}
\label{matriz_worthey}
\end{equation}
being $S_{M_l}$ the metal sensitivity parameter introduced by Worthey (1994)
for the $l^{\rm th}$ line-strength index $M_l$. It is important to note that
the method followed by Worthey (1994, see his footnote number 4) is a good
approximation for the derivatives $d\log({\rm age})/d\log({\rm Z})$ as
long as the approximation that the $a_{ij}$ coefficients are constant in the
region where age and metallicity are being measured holds. However, since this
is not always the case, in Appendix~\ref{appendix_sensitivity} we describe the
use of bivariate polynomials to express locally the $a_{ij}$ coefficients as a
function of age and metallicity, which in turn can be employed to derive more
accurate local metal sensitivity parameters through the relations given in
Eq.~(\ref{matriz_worthey}).

If the $2\times2$ matrix $A$ of coefficients $a_{ij}$ in
Eq.~(\ref{eq_linear_approx}) is invertible, one can also express locally age
and metallicity variations as a function of variations in the line-strength
indices by
\begin{equation}
\left(\begin{array}{@{}c@{}} \Delta\log(Z) \\ 
                             \Delta\log({\rm age}) \end{array}\right) =
\left(\begin{array}{@{}c@{\;\;}c@{}} b_{11} & b_{12} \\
                                     b_{21} & b_{22} \\ \end{array}\right)
\left(\begin{array}{@{}c@{}} \Delta m_1 \\ \Delta m_2 \end{array}\right),
\label{error_matriz}
\end{equation}
where
\begin{equation}
\begin{array}{@{}lc@{}l}
b_{11} = & & a_{22}/\det(A),
\\ 
b_{12} = &-& a_{12}/\det(A),
\\ 
b_{21} = &-& a_{21}/\det(A),
\\ 
b_{22} = & & a_{11}/\det(A).
\end{array}
\label{equation_b_from_a}
\end{equation}

Assuming that the indices $M_1$ and $M_2$ are independent, replacing $\Delta
m_l$ in Eq.~(\ref{error_matriz}) by the random error $\sigma(m_l)$, and using
the result given in Eq.~(\ref{equation_error_sn}), the area of the
2-dimensional error-ellipse in the space defined by the physical parameters
$\log(Z)$ and $\log({\rm age})$, can be computed as
\begin{equation}
\begin{array}{l}
V_{\log(Z),\log({\rm age})} = \\ \noalign{\medskip}
\;\;\;\;  \pi \; \chi^2_2(\alpha) \; | \det(B) | \; 
            \displaystyle
              \frac{c(m_1)}{{\it SN}(\mbox{\AA})_1} \;
              \frac{c(m_2)}{{\it SN}(\mbox{\AA})_2} = \\ \noalign{\medskip}
\;\;\;\;  \pi \; \chi^2_2(\alpha) \;
            \displaystyle\frac{1}{| \det(A) |} \; 
            \displaystyle
              \frac{c(m_1)}{{\it SN}(\mbox{\AA})_1} \;
              \frac{c(m_2)}{{\it SN}(\mbox{\AA})_2},
\end{array}
\end{equation}
from which it is evident that the suitability parameter is defined as
\begin{equation}
\kappa \equiv | \det(B) | \; c(m_1) \; c(m_2) =
\underbrace{\displaystyle\frac{1}{| \det(A) |}}_{
    \begin{array}{c} 
      \mbox{\small sensitivity to} \\ 
      \mbox{\small age--metallicity} \\ 
      \mbox{\small degeneracy}
    \end{array}}
\; 
\underbrace{c(m_1) \; c(m_2)}_{
    \begin{array}{c} 
      \mbox{\small sensitivity} \\ 
      \mbox{\small to {\it SN}(\mbox{\AA})} 
    \end{array}},
\label{equation_kappa}
\end{equation}
where the effect of the degeneracy of the iso-metallicity and
iso-age model predictions are encapsulated in the geometric transformation
represented by the $a_{ij}$ coefficients (or by the $b_{ij}$ coefficients of
the inverse transformation), whereas the sensitivity to ${\it SN}(\mbox{\AA})$
is controlled by the $c(m_l)$ factors. 

Thus, it is clear that the goal of achieving minimum total errors in both age
and metallicity involve the balancing of the two effects. In fact, degeneracy
and sensitivity to errors typically behave in opposite sense as a function of
the wavelength coverage of the considered spectroscopic feature. Narrow
line-strength indices can be found to be sensitive almost only to age or to
metallicity, whereas broad band features are less sensitive to noise, as can be
seen in the functional dependence of $c(m_l)$ on bandpass widths
(Eq.~(\ref{coefficients_vs_bandpass_width}) and Eq.~(\ref{equation_xi1xi2})
for line-strength indices and colors, respectively).

\renewcommand\arraystretch{1.3}
\begin{table*}[t]
\center{\scriptsize
\input{rmodel1.inc}

\caption{Logarithm of the suitability index, $\log[\kappa]$, ---see
Eq.~(\ref{equation_kappa})--- for the study of the age-metallicity degeneracy,
computed for different combinations of common line-strength indices in the
optical range.  For the generation of these numbers we have employed the
predictions of Bruzual \& Charlot (2001) models, for single stellar populations
of 12~Gyr old and solar metallicity. Better diagrams are those for which
$\kappa$ are lower. In this sense, and as a guide for the eye, we have
boldfaced and underlined the 10\% of these numbers with the lowest values.}
\label{table_suitab_indice_indice}
}
\end{table*}
\renewcommand\arraystretch{1.0}

\renewcommand\arraystretch{1.3}
\begin{table*}[t]
\center{\scriptsize
\input{rmodel2.inc}

\caption{Same than Table~\ref{table_suitab_indice_indice} but for
different combinations of colors.}
\label{table_suitab_color_color}
}
\end{table*}
\renewcommand\arraystretch{1.0}

\renewcommand\arraystretch{1.3}
\begin{table*}[t]
\center{\scriptsize
\input{rmodel3.inc}

\caption{Same than Table~\ref{table_suitab_indice_indice} but for
different combinations of line-strength indices and colors.}
\label{table_suitab_indice_color}
}
\end{table*}
\renewcommand\arraystretch{1.0}

In Tables~\ref{table_suitab_indice_indice}--\ref{table_suitab_indice_color}
we present suitability parameters (more precisely $\log[\kappa]$) computed for
different combinations of common line-strength indices and colors. In their
computation we have employed the predictions of simple stellar populations from
Bruzual \& Charlot (2001)\footnote{Available at\\
http://www.sdss.mpg.de/sdssMPA/Spectral\_Tools/}. In addition, in
Table~\ref{table_suitab_additional_indices} we have also determined the
suitability parameter for those combinations of line-strength indices including
narrow features, that so far have been considered to provide a better
discrimination between age and metallicity: Fe4668 as metallicity indicator,
and H$\gamma_{\rm HR}$ (Jones \& Worthey~\cite{jones95}), H$\delta_{\rm A}$,
H$\delta_{\rm F}$, H$\gamma_{\rm A}$ and H$\gamma_{\rm F}$ (Worthey \&
Ottaviani~\cite{worthey97}), and H$\gamma_{\mbox{\scriptsize VA},\sigma}$
(Vazdekis \& Arimoto~\cite{vazdekis99}) as age 
indicators\footnote{In the computation of
the numbers listed in
Tables~\ref{table_suitab_indice_indice}--\ref{table_suitab_indice_indice_comp}
we have assumed that all the spectral data are not correlated. Since some of
the line-strength indices and colors overlap, we are implicitly assuming that
overlapping features are obtained in independent measurements (i.e.\ different
spectra and photometric colors).}.

\begin{table}[t]
\centering
\begin{tabular}{ccc} \hline
index-index diagram            &  $\log[\kappa]$ & models   \\ \hline
H$\gamma_{\rm HR}$ vs.\ Fe4668 &    1.3          & W94,JW95 \\
H$\delta_{\rm A}$ vs.\ Fe4668  &    0.7          & W94,WO96 \\
H$\delta_{\rm F}$ vs.\ Fe4668  &    0.9          & W94,WO96 \\
H$\gamma_{\rm A}$ vs.\ Fe4668  &    0.7          & W94,WO96 \\
H$\gamma_{\rm F}$ vs.\ Fe4668  &    0.8          & W94,WO96 \\
H$\gamma_{\mbox{\scriptsize VA,125}}$ vs.\ Fe4668
                               &    1.5          & VA99  \\
H$\gamma_{\mbox{\scriptsize VA,200}}$ vs.\ Fe4668
                               &    1.7          & VA99  \\
H$\gamma_{\mbox{\scriptsize VA,275}}$ vs.\ Fe4668
                               &    1.8          & VA99  \\ \hline
\end{tabular}
\caption{H$\gamma_{\mbox{\scriptsize VA},\sigma}$ correspond to the
$[\mbox{H}\gamma+1/2(\mbox{Fe~{\sc i}}+\mbox{Mg~{\sc i}})]_\sigma$ indices
defined by Vazdekis \& Arimoto~(\cite{vazdekis99}).  Model references are W94:
Worthey~(\cite{worthey94}); JW95: Jones \& Worthey~(\cite{jones95}); WO97:
Worthey \& Ottaviani~(\cite{worthey97}); and V99: Vazdekis \&
Arimoto~(\cite{vazdekis99}).}
\label{table_suitab_additional_indices}
\end{table}

\subsection{Two examples}
\label{section_two_examples}

\begin{figure*}
 \resizebox{1.0\hsize}{!}{\includegraphics[bb=33 278 545 514]{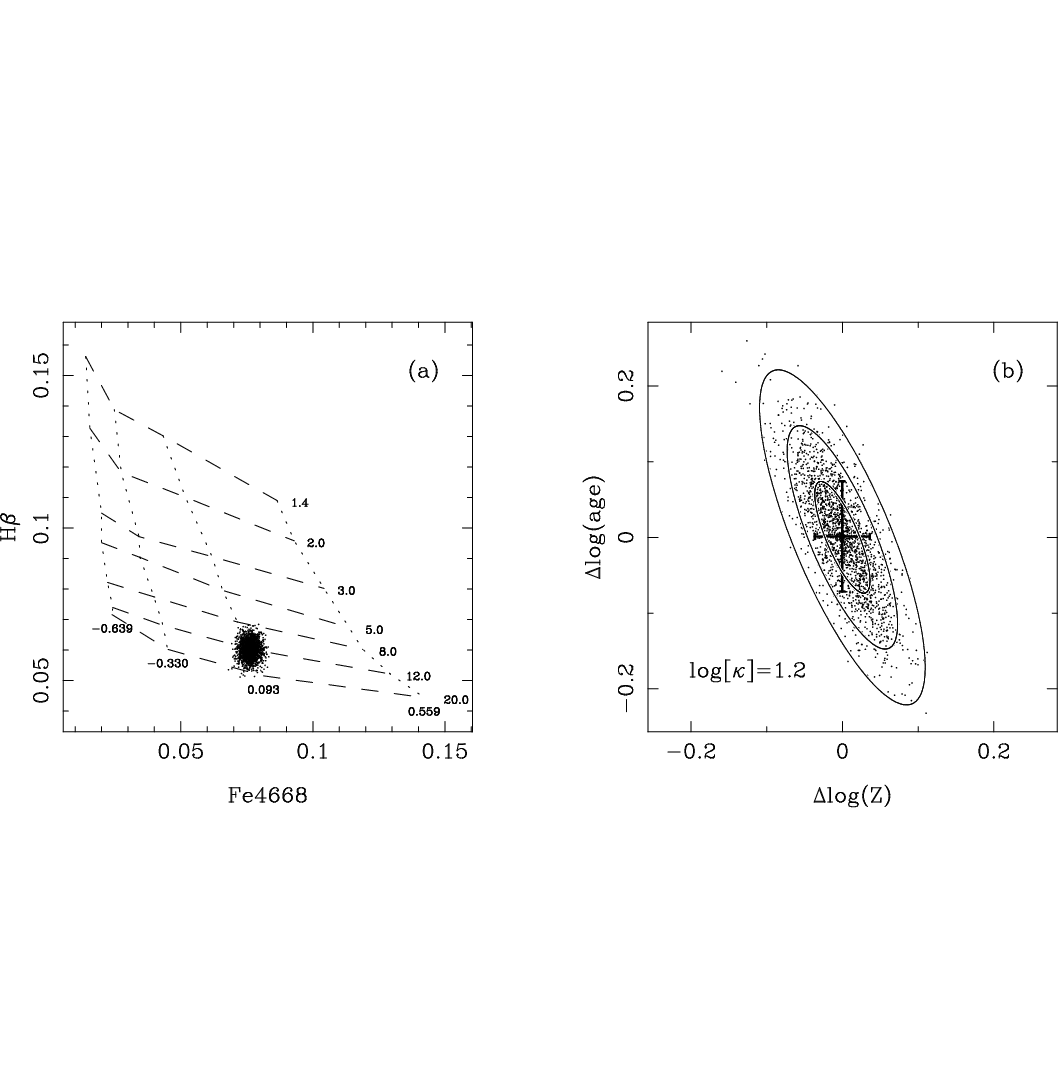}}

 \vspace{7mm}

 \resizebox{1.0\hsize}{!}{\includegraphics[bb=33 278 545 514]{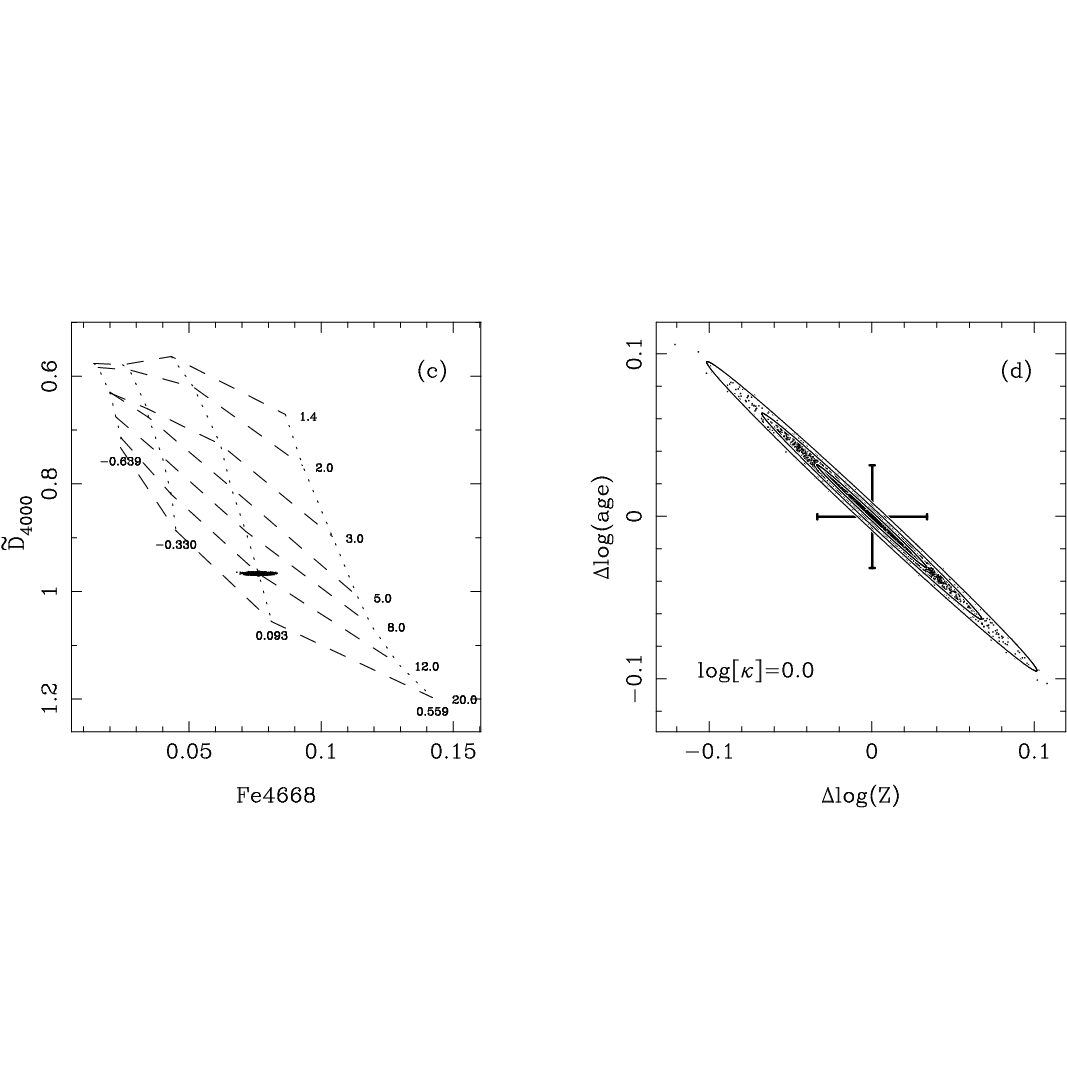}}
 \caption{{\bf Panel~(a)}: H$\beta$--Fe4668 diagram, with Bruzual \&
 Charlot~(2001) models over-plotted (dashed lines and dotted lines correspond to
 the predictions for constant age and metallicity, respectively; ages are
 labeled in Gyr, and metallicities are given as [Fe/H]). Dots correspond
 to 2000 simulations of a 12 Gyr old object with solar metallicity,
 assuming $\mbox{\rm {\it SN}({\rm \AA})} = 100$. {\bf Panel~(b)}: Errors in
 $\log({\rm age})$ and $\log(Z)$ for the same simulated data displayed in
 panel~(a). Ellipses indicate the regions of 68.26, 95.44 and 99.73\%
 probability, whereas the central error bars show the unbiased standard
 deviation in each axis. {\bf Panels~(c) and~(d)}: Same than panels~(a) and~(b)
 after introducing the $\tilde{\rm D}_{4000}$ index instead of H$\beta$, and
 using the same $\mbox{\rm {\it SN}({\rm \AA})}$ in the simulations. The
 $\tilde{\rm D}_{4000}$ index is plotted in an inverse scale to display the
 model predictions with roughly the same orientation in age and metallicity
 than in panel~(a). Note also that the axis scales in panels~(b) and~(d) are
 different. See discussion in Sect.~\ref{section_two_examples}.}
 \label{plot_pds}
\end{figure*}

\begin{table}[t]
\centering
\begin{tabular}{lr|ccc} \hline
\multicolumn{5}{c}{Line-strength indices from BC01}    \\ \hline
         &            & \multicolumn{3}{c}{[Fe/H]}     \\
         &            & $-0.330$ &  $0.093$ &  $0.059$ \\ \hline
         &      8 Gyr & 0.749313 & 0.888150 & 1.062612 \\
$\tilde{\rm D}_{4000}$ &
               12 Gyr & 0.819419 &\bf 0.966693 & 1.135206 \\
         &     20 Gyr & 0.887191 & 1.056064 & 1.197137 \\ \hline
         &      8 Gyr & 0.077433 & 0.069191 & 0.060212 \\
H$\beta$ &     12 Gyr & 0.068105 &\bf 0.060252 & 0.052139 \\
         &     20 Gyr & 0.060212 & 0.051545 & 0.044401 \\ \hline
         &      8 Gyr & 0.039794 & 0.071434 & 0.118949 \\
Fe4668   &     12 Gyr & 0.043234 &\bf 0.076271 & 0.129336 \\
         &     20 Gyr & 0.045004 & 0.081293 & 0.142160 \\ \hline
\multicolumn{5}{c}{ }                                  \\
\end{tabular}
\begin{tabular}{lrrc} \hline
         & \multicolumn{2}{c}{diagram}         &                Eq. \\ 
parameter& H$\beta$--Fe4668 & $\tilde{\rm D}_{4000}$--Fe4668  &reference \\ 
\hline
$p_{10}$ & $-0.430095$      & $-0.430095$      &(\ref{matrixA_bivariate})\\
$p_{11}$ & $ 0.051432$      & $ 0.051432$      &(\ref{matrixA_bivariate})\\
$p_{20}$ & $ 0.040343$      & $ 0.040343$      &(\ref{matrixA_bivariate})\\
$p_{01}$ & $ 0.323917$      & $ 0.323917$      &(\ref{matrixA_bivariate})\\
$p_{02}$ & $-0.014719$      & $-0.014719$      &(\ref{matrixA_bivariate})\\
$q_{10}$ & $-0.058764$      & $ 0.488931$      &(\ref{matrixA_bivariate})\\
$q_{11}$ & $ 0.003947$      & $-0.013260$      &(\ref{matrixA_bivariate})\\
$q_{20}$ & $ 0.002449$      & $-0.013883$      &(\ref{matrixA_bivariate})\\
$q_{01}$ & $-0.717729$      & $ 4.921280$      &(\ref{matrixA_bivariate})\\
$q_{02}$ & $ 0.033402$      & $-0.225252$      &(\ref{matrixA_bivariate})\\
log[age] & $10.079182$      & $10.079182$      &(\ref{matrixA_bivariate})\\
log[Z]   & $ 0.093200$      & $ 0.093200$      &(\ref{matrixA_bivariate})\\
\hline
$a_{11}$ & $ 0.095814$      & $ 0.095814$      & 
                      (\ref{eq_linear_approx}), (\ref{matrixA_bivariate})\\
$a_{12}$ & $ 0.031999$      & $ 0.031999$      &
                      (\ref{eq_linear_approx}), (\ref{matrixA_bivariate})\\
$a_{21}$ & $-0.018528$      & $ 0.352696$      &
                      (\ref{eq_linear_approx}), (\ref{matrixA_bivariate})\\
$a_{22}$ & $-0.044036$      & $ 0.379328$      &
                      (\ref{eq_linear_approx}), (\ref{matrixA_bivariate})\\
\hline
$c(m_1)$ & 0.2757           & 0.1086           &(\ref{equation_kappa})   \\
$c(m_2)$ & 0.2235           & 0.2235           &(\ref{equation_kappa})   \\
$\kappa$ & 16.995           & 0.9685           &(\ref{equation_kappa})   \\
\hline
\end{tabular}
\caption{Numerical values for the example of Section~\ref{section_two_examples}
(see also Fig.\ref{plot_pds}). The upper table lists the input line-strength
indices employed in the computation of the bivariate polynomial transformations
in Figs.~\ref{plot_pds}(a) and~\ref{plot_pds}(c) around the model predictions
for a SSP of 12~Gyr with solar metallicity (boldfaced). The lower table 
displays the different parameters involved in the calculation of the previous
transformations, and the factors leading to the {\it suitability index\/} in
Eq.~(\ref{equation_kappa}).}

\label{table_numbers_from_example}
\end{table}

In Fig.~\ref{plot_pds}, panels~(a) and~(c), we show two examples of typical
index-index diagrams. In both cases we have simulated the effect of observing a
hypothetical simple stellar population 12~Gyr old and with [Fe/H]=0.093
(corresponding to the physical parameters of one of the points predicted by the
grid of models), with a signal-to-noise ratio per \AA\ of 100. The simulations
are displayed as clouds of points clustered around the model prediction for
those physical parameters. Each simulated point in the index-index diagrams has
been transformed into age and metallicity using a $N=2$ bivariate polynomial
transformation, as explained in Appendix~\ref{appendix_sensitivity} (see
Fig.~\ref{diagrama_Hbeta_Fe4668}). As a guide to the reader, we give in
Table~\ref{table_numbers_from_example} the numbers involved in the computation
of the suitability parameter in the two index--index diagrams shown in
Fig.~\ref{plot_pds}. The difference between the physical parameters derived
from each simulated point, relative to the values corresponding to the
hypothetical stellar population, are represented as errors in $\log({\rm age})$
and $\log(Z)$ in panels~(b) and~(d). The ellipses show the regions of 68.26,
95.44 and 99.73\% probability.  The areas covered by these ellipses in the
error space are clearly larger in the H$\beta$-Fe4668 diagram than in the
$\tilde{\rm D}_{4000}$-Fe4668 diagram (note that the axis scales in panels~(b)
and~(d) are different). This result agrees with the larger (and thus worse)
value of $\log[\kappa]$ for the first index-index diagram. Note, however, that
the errors are more correlated in the second diagram. In fact, the standard
deviations around each physical parameter, displayed with thick error bars, are
a relatively fair representation of the $1\,\sigma$ error ellipse in panel~(b),
but not in panel~(d). For that reason, the standard deviations by themselves
are not a good parametrization of the actual uncertainty in the derived
parameters.

In practice, when one tries to answer the question of whether the integrated
light of two stellar populations share the same underlying physical parameters
{\it within the error bars\/}, the question translates into whether their error
ellipses overlap in the space defined by those physical parameters.  Since the
probability of overlapping decreases as the area of the error ellipses becomes
smaller, the {\it suitability parameter\/} is a direct indication of such
probability. The presence of correlation between the errors is not critical, as
far as the fake relationship, introduced by the presence of error correlation,
is taken into account when studying measurements performed in different
objects. In this sense, the use of numerical simulations may help to analyze
the relative contribution of such error correlations to the intrinsic
relationships between the physical parameters (e.g.\ Kuntschner et
al.~\cite{kuntschner01}).

\subsection{Selecting the most suitable diagram}
\label{section_selecting_diagram}

It is clear from the previous discussion that the best \mbox{$M_1$--$M_2$}
diagram to
disentangle physical parameters (and, in particular, age and metallicity) will
be that for which the factor
\begin{equation}
\psi_{M_1,M_2} = \kappa \;
  \displaystyle\frac{1}{{\it SN}(\mbox{\AA})_1} \;
  \displaystyle\frac{1}{{\it SN}(\mbox{\AA})_2}
\label{factor_a_minimizar}
\end{equation}
is minimum. Although initially this can be achieved by selecting the
combination of spectral measurements (indices and colors) for which
$\log[\kappa]$ is lowest, in practice one should consider realistic
signal-to-noise ratios. For that reason, and although at the light of the
results displayed in
Tables~\ref{table_suitab_indice_indice}--\ref{table_suitab_additional_indices}
it seems that color-color diagrams are the best option, in practice this is not
necessarily the case. For instance, typical random errors in the measurement of
colors are of the order of 0.01~mag (e.g.\ Babu \& Feigelson \cite{babu96}),
which implies signal-to-noise ratios per \AA\ of the order of a few (see
Fig.~\ref{plot_errsn}). However, spectroscopic line-strength indices are
commonly obtained with ${\it SN}(\mbox{\AA})\ga 10$ times larger. 

In order to compare the suitability of diagrams built with two line-strength
indices, two colors, or one line-strength index and one color, let assume that
\begin{equation}
{\it SN}(\mbox{\AA})_{\rm line-strength\;index} \simeq 
   10 \; {\it SN}(\mbox{\AA})_{\rm color}.
\label{hipotesis_ruido} 
\end{equation}
Under this hypothesis, Eq.~(\ref{factor_a_minimizar}) provides
\begin{equation}
\psi_{\rm color_1,color_2} \simeq 100 \; 
 \frac{ \kappa_{\rm color_1,color_2}}{\kappa_{\rm index_1,index_2}} \;
 \psi_{\rm index_1,index_2},
\end{equation}
and
\begin{equation}
\psi_{\rm index,color} \simeq 10 \; 
 \frac{ \kappa_{\rm index,color}}{\kappa_{\rm index_1,index_2}} \;
 \psi_{\rm index_1,index_2}.
\end{equation}

Considering that the values displayed in
Tables~\ref{table_suitab_indice_indice}--\ref{table_suitab_additional_indices}
correspond to $\log[\kappa]$, and focusing in the best values (the ones
highlighted and underlined in these tables), we see that
\begin{equation}
 \frac{ \kappa_{\rm color-color}}{\kappa_{\rm index-index}} \simeq 0.01,
\end{equation}
and
\begin{equation}
 \frac{ \kappa_{\rm index-color}}{\kappa_{\rm index-index}} \simeq 0.1.
\end{equation}
Finally, combining all these numbers we obtain
\begin{equation}
\psi_{\rm color_1,color_2}^{\rm best} \simeq
\psi_{\rm index_1,color_2}^{\rm best} \simeq
\psi_{\rm index_1,index_2}^{\rm best}.
\end{equation}
This result, derived after assuming the crude hypothesis stated in
Eq.~(\ref{hipotesis_ruido}), indicates that {\it the best\/} index-index,
index-color and color-color diagrams are, initially, almost equally suited for
the study of the age-metallicity degeneracy (in 12~Gyr-old simple stellar
populations with solar metallicity). In a real situation, the selection
of line-strength features, colors, or a combination of both to build a good
diagram will depend on the available signal-to-noise ratio for each spectral
indicator.

Focusing on the results displayed in
Tables~\ref{table_suitab_indice_indice}--\ref{table_suitab_indice_color}, and
leaving in a second plane the relevance of the signal-to-noise ratio just
discussed, the $\tilde{\rm D}_{4000}$, Fe4668 and Mg$_2$ features are the best
line-strength features to be included in index-index and index-color diagrams,
whereas for color-color diagrams the lowest $\log[\kappa]$ values are obtained
for colors involving well separated bandpasses, like $(g-K_s)$, and $(V-K)$.

Interestingly, the suitability parameter for combinations of narrow
line-strength features with Fe4668 (displayed in
Table~\ref{table_suitab_additional_indices}) are worse than the value for the
$\tilde{\rm D}_{4000}$-Fe4468 diagram, and only $H\delta_{\rm A}$,
$H\delta_{\rm F}$, $H\gamma_{\rm A}$, and $H\gamma_{\rm F}$ can rival with
Mg$_2$. If, in addition, we consider that the $c(M)$ coefficients
(Table~\ref{table_c_coefficients}) for the narrow indices are larger than the
same coefficients for the broader features $\tilde{\rm D}_{4000}$ and Mg$_2$,
at a fixed signal-to-noise ratio per \AA\ the diagrams of these two latest
spectral features with Fe4668 provide more information.

\subsection{Some words of caution}
\label{section_problemas}

It is very important to keep in mind that, for several reasons, the above
results must be taken with care:

(i) The list of line-strength indices and colors explored is, obviously, not
complete.

(ii) The derived numbers do not take into account the uncertainties in the
stellar population modeling. In fact, in
Table~\ref{table_suitab_indice_indice_comp} we compare the suitability
parameters derived using the predictions of three different sets of models. The
differences exhibited by some combinations of line-strength indices reflect
the existing discrepancies between different models. However, it is worth
noting that for those combinations involving Fe4668 the agreement is very
reasonable, and less good for $\tilde{\rm D}_{4000}$, Mg$_2$, and Na5895.

(iii) The integrated light of the stellar systems under study (e.g.\ star
clusters, galaxies) are not necessarily well described by simple stellar
populations. New models, including more complex star formation histories, may
provide a different suitability ranking of line-strength features and colors.
Anyway, the procedure here described is still valid as long as the correct
models are employed.

\begin{figure}
 \resizebox{1.0\hsize}{!}{\includegraphics[angle=-90]{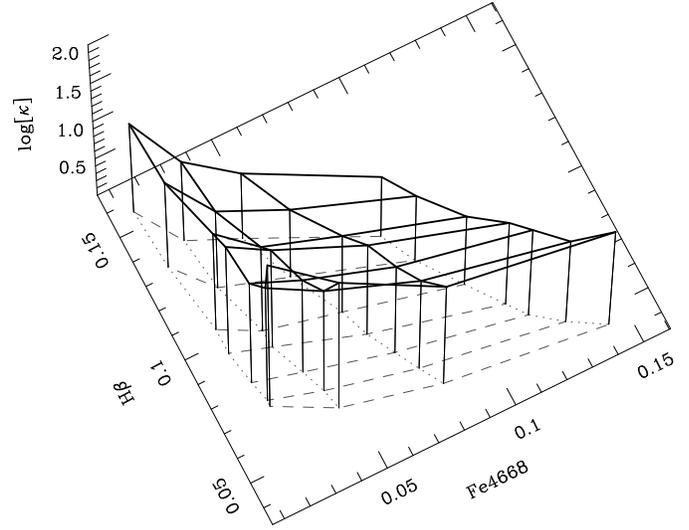}}
 \caption{Three-dimensional representation of the $\log[\kappa]$ values for all
 the intersecting points in the grid of models from Bruzual \& Charlot (2001),
 for the H$\beta$--Fe4668 diagram.}
 \label{plot_project3D}
\end{figure}

(iv) When comparing several objects in a diagram, their physical parameters are
expected to exhibit a range.  Since the $\kappa$ values shown in
Tables~\ref{table_suitab_indice_indice}--\ref{table_suitab_additional_indices}
have been computed for a fixed age and metallicity, the suitability parameter
will be different for each object. For illustration, in
Fig.~\ref{plot_project3D} we represent a three-dimensional plot with the
$\log[\kappa]$ values derived for each intersecting point constituting the grid
of models in the H$\beta$--Fe4668 diagram. It is clear from this figure that
the suitability parameter is a function of the location in the index-index
plane.

(v) Although systematic effects have not been considered in the analysis
performed so far (see Section~\ref{sect_error_handling}), in the real world
they can be the ones that, if not properly constrained, may determine the
actual suitability of a given set of spectroscopic observables.  In practice
the nature and relevance of systematic errors affecting line-strength indices
(e.g.\ spectral resolution, radial velocity, flux calibration) and colors
(e.g.\ zeropoint ca\-li\-bra\-tion, \mbox{$k$-corrections}) are typically
different.  For that reason, and assuming that the spectroscopic data (either
spectra or colors, but not both) are obtained under homogeneous conditions, it
is safer to use the numbers displayed in
Tables~\ref{table_suitab_indice_indice} and~\ref{table_suitab_color_color} in a
differential way, i.e.\ as a guide to compare different diagrams within each of
these tables, but not to compare the absolute values of the suitability indices
between the two tables. The same reasoning makes the numbers displayed in
Table~\ref{table_suitab_indice_color} more uncertain that the ones shown in
Tables~\ref{table_suitab_indice_indice} and~\ref{table_suitab_color_color}.

Summarizing, and even considering all these problems, it is clear that once a
given set of models has been adopted, the factor displayed in
Eq.~(\ref{factor_a_minimizar}) is an excellent tool to estimate an optimized
combination of spectroscopic measurements, in order to face the study of
integrated spectra.

\renewcommand\arraystretch{1.3}
\begin{table*}[t]
\center{\scriptsize
\input{rmodel_comparison.inc_modified}

\caption{Comparison of $\log[\kappa]$ for the study of the age-metallicity
degeneracy, computed from the prediction of three sets of models, W94
(Worthey~\cite{worthey94}), V00 (Vazdekis et al.~\cite{vazdekis96}; Blakeslee
et al.~\cite{blakeslee01}), and BC01 (Bruzual \& Charlot~2001). In all the
cases we have derived the suitability parameters for SSP around 12~Gyr old and
solar metallicity.}
\label{table_suitab_indice_indice_comp}
}
\end{table*}
\renewcommand\arraystretch{1.0}

%
\section{Conclusions}
\label{section_conclusions}

We have investigated the combined role of parameter degeneracy and
signal-to-noise ratio in the study of the integrated spectroscopic properties
of astronomical objects. In particular, we have examined the effect of random
errors at the light of stellar population model predictions. We have shown that
the expected random error in the measurement of line-strength indices and
colors can be very easily computed as a constant divided by an appropriate
average of the signal-to-noise ratio per \AA, as stated in
Eq.~(\ref{equation_error_sn}). This simple expression allows to define a
suitability parameter which combines both effects (degeneracy and sensitivity
to noise), providing an immediate tool to compare the usefulness of different
observational diagrams. The recipe to perform such a comparison is the
following:

(i) Chose a reliable stellar population model.

(ii) Obtain the geometric transformation to convert the observational
parameters (line-strength indices and colors) into physical data (age,
metallicity, initial mass function,\ldots). For this purpose, bivariate
polynomial transformations (Appendix~\ref{appendix_sensitivity}) are a very
convenient way to obtain the coefficients given in
Eq.~(\ref{matrixA_bivariate}), or those corresponding to the inverse
transformation, Eq.~(\ref{equation_b_from_a}).

(iii) Use Eq.~(\ref{equation_kappa}) to obtain $\kappa$. The coefficients which
indicate the sensitivity of the spectroscopic data to the signal-to-noise ratio
can be extracted from Table~\ref{table_c_coefficients}.

(iv) Using reasonable estimates of the expected signal-to-noise ratio per \AA\
in each spectroscopic measurement, employ Eq.~(\ref{factor_a_minimizar}) to
obtain $\psi$. As discussed in Sect.~\ref{section_selecting_diagram}, this is
the factor to be minimized.

Classical atomic line-strength indices must be measured as molecular indices in
order to apply the above procedure. The same also holds for generic indices
(e.g.\ CaT, PaT, CaT$^{*}$; see Cenarro et al.~\cite{cenarro01}).

We have illustrated this method by studying in more detail the well known
age--metallicity degeneracy. Using model predictions for a 12~Gyr old simple
stellar population with solar metallicity, we have shown that a broad spectral
feature like the D$_{4000}$ can be as well suited (or even more) than H$\beta$
to analyze this kind of degeneracy, once the dependence on signal-to-noise
ratio is taken into account.

For all the reasons mentioned in Sect.~\ref{section_problemas}, the aim of this
paper is not to give a definite answer to the question of which is the best
observational space to disentangle physical properties of stellar populations,
but to provide an easy way to determine the relative suitability of different
spectroscopic diagrams to obtain physical information of the astronomical
objects under study.

%
\begin{acknowledgements}
Valuable discussions with A.\ Vazdekis, and J.\ J.\ Gonz\'{a}lez are gratefully
acknowledged.  The data employed in Appendix~\ref{appendix_d4000} was obtained
with the 1m~JKT, 2.5m~INT and the 4.2m~WHT at La Palma Observatory, and with
the 3.5m Telescope at Calar Alto Observatory. The JKT, INT and WHT are operated
on the island of La Palma by the Royal Greenwich Observatory at the
Observatorio del Roque de los Muchachos of the Instituto de Astrof\'{\i}sica de
Canarias. The Calar Alto Observatory is operated jointly by the
Max-Planck-Institute f\"{u}r Astronomie, Heidelberg, and the Spanish
Comisi\'{o}n Nacional de Astronom\'{\i}a.  This work was supported by the
Spanish Programa Nacional de Astronom\'{\i}a y Astrof\'{\i}sica under grant
AYA2000-977.
\end{acknowledgements}

%

%
\appendix

\section{Estimating random errors in the D$_{4000}$}
\label{appendix_d4000}

In order to explore in more detail the validity of
Eq.~(\ref{equation_error_sn_old_d4000}), we have compared the error estimations
in the D$_{4000}$ derived from that formulae with the results derived by
using a more accurate method.

\begin{figure}
  \resizebox{1.0\hsize}{!}{%
  \includegraphics[angle=0]{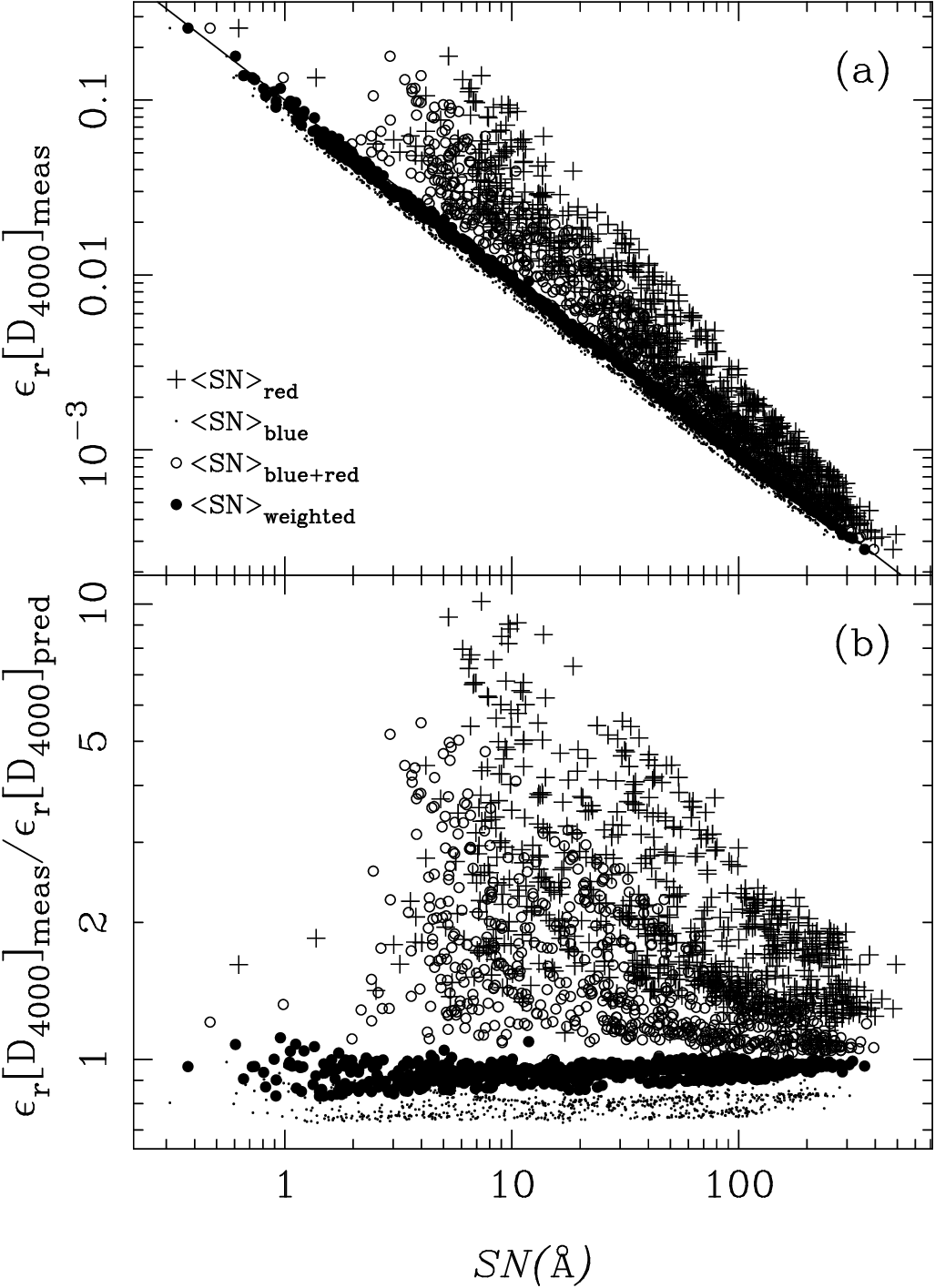}}
  \caption{Relative errors in the D$_{4000}$, $\epsilon_{\rm r}[{\rm
  D}_{4000}]$, as a function of the signal-to-noise per \AA, measured in the
  stellar library employed by Gorgas et al.~(\cite{gorgas99}) to derive the
  empirical calibration of this spectral feature.  See text in
  Appendix~\ref{appendix_d4000} for details.}
\label{fig_errd4000_vs_sn}
\end{figure}

In Fig.~\ref{fig_errd4000_vs_sn} we represent the relative errors in the
D$_{4000}$, $\epsilon_{\rm r}[{\rm D}_{4000}]$, as a function of the
signal-to-noise per \AA, as measured in the 713 spectra (including repeated
observations) of the stellar library gathered by Gorgas et
al.~(\cite{gorgas99}) to derive the empirical calibration of this spectral
feature. In panel~(a) we present the measured relative error determined using
an accurate method ---Eqs.~(38)--(40) of Cardiel et
al.~(\cite{cardiel98})\mbox{---,}
for four different estimations of the signal-to-noise ratio per \AA.  In
particular, crosses and small dots indicate the results obtained when using the
mean ${{\it SN}(\mbox{\AA})}$ in the blue and red bandpasses, respectively; the
signal-to-noise ratio for the open circles has been computed as the arithmetic
mean of the two previous values, whereas for the filled circles we have
employed the weighted mean $(1/\langle{\it SN}\rangle_{\rm blue}+1/\langle{\it
SN}\rangle_{\rm red})/ (1/\langle{\it SN}\rangle^2_{\rm blue}+1/\langle{\it
SN}\rangle^2_{\rm red})$. The prediction of
Eq.~(\ref{equation_error_sn_old_d4000}) is the diagonal full line, whereas the
residuals with respect to this prediction are plotted in panel~(b).  The
employed stellar sample typically contains spectra with poorer signal-to-noise
ratio in the blue bandpass of the D$_{4000}$ than in the red bandpass, and the
simple arithmetic mean of the ${{\it SN}(\mbox{\AA})}$ is not a good
approximation.  

It is clear from the previous figure that the weighted mean is the best
approximation. We have also checked that a weighted mean of the form
$[2/(1/\langle{\it SN}\rangle^2_{\rm blue}+1/\langle{\it SN}\rangle^2_{\rm
red})]^{1/2}$ (not shown in the figure) gives acceptable results.

\section{Estimating random errors in colors}
\label{appendix_colors}

With the aim of estimating the dependence of random errors in colors on the
signal-to-noise ratio, we are following here a similar procedure to that
employed in Cardiel et al.~(\cite{cardiel98}) to derive the corresponding
formulae for classical indices. 

Given two filters, a spectral energy
distribution $f(\lambda)$, and the SED of a reference object ${\cal
F}(\lambda)$ (e.g.\ the SED of $\alpha$Lyr for magnitudes measured in the Vega
system, ${\cal F}(\nu)=3.63\times 10^{-20}$~erg~cm$^{-2}$~s$^{-1}$~Hz$^{-1}$
for AB magnitudes, or ${\cal F}(\lambda)=3.63\times
10^{9}$~erg~cm$^{-2}$~s$^{-1}$~\AA$^{-1}$ for HST magnitudes), a color can
be determined by (see e.g.\ Fukugita et al.~\cite{fukugita95})
\begin{equation}
\begin{array}{@{}ll}
C = & -2.5\;\log_{10} \left[
\frac{\displaystyle\int_\lambda f(\lambda) \; R_1(\lambda) \; 
{\rm d}\lambda}{\displaystyle\int_\lambda f(\lambda) \; R_2(\lambda) \; 
{\rm d}\lambda} \right] \\ \noalign{\medskip}
    & +2.5\;\log_{10} \left[
\frac{\displaystyle\int_\lambda {\cal F}(\lambda) \; R_1(\lambda)\;
{\rm d}\lambda}{\displaystyle\int_\lambda {\cal F}(\lambda) \; R_2(\lambda)\; 
{\rm d}\lambda} \right],
\end{array}
\end{equation}
where $R_i(\lambda)$ is the response function of the $i^{\rm th}$~filter.
The previous expression can be rewritten as
\begin{equation}
C = -2.5\;\log_{10} \frac{\langle f(\lambda) \rangle_1}%
                         {\langle f(\lambda) \rangle_2}
    \; + C_0,
\label{equation_color_meanfluxes}
\end{equation}
where
\begin{equation}
\langle f(\lambda) \rangle_i = 
  \frac{\displaystyle\int_\lambda f(\lambda)\;R_i(\lambda)\;{\rm d}\lambda}%
{\displaystyle\int_\lambda R_i(\lambda) \; {\rm d}\lambda},
\label{equation_meanflux}
\end{equation}
\begin{equation}
  C_0=-2.5\;\log_{10} \frac{k_1}{k_2} \; + \tilde{C_0},
\end{equation}
\begin{equation}
k_i = \displaystyle\int_\lambda R_i(\lambda) \; {\rm d}\lambda
\label{equation_ki}
\end{equation}
and
\begin{equation}
\tilde{C_0} =+2.5\;\log_{10} \left[
\frac{\displaystyle\int_\lambda {\cal F}(\lambda) \; R_1(\lambda)\;
{\rm d}\lambda}{\displaystyle\int_\lambda {\cal F}(\lambda) \; R_2(\lambda)\; 
{\rm d}\lambda} \right].
\end{equation}
In practice integrals must be replaced by summations of the form
\begin{equation}
\displaystyle\int_\lambda f(\lambda)\;R_i(\lambda)\;{\rm d}\lambda \simeq
\Theta\;\displaystyle\sum_{k=1}^{N_i} f(\lambda_{i,k})\; R_i(\lambda_{i,k}),
\end{equation}
where $\Theta$ is the linear dispersion (in \AA~pixel$^{-1}$, assumed to be
constant along the spectrum), $N_i$ is the number of pixels covering the
$i^{\rm th}$ filter, $\lambda_{i,k}$ is the central wavelength of the
$k^{\rm th}$ pixel of the $i^{\rm th}$ filter, and
$f(\lambda_{i,k})$ is the flux is that pixel.

If $\sigma_f(\lambda)$ is the error spectrum associated to $f(\lambda)$, and if
we assume that ${\cal F}(\lambda)$ and $R_i(\lambda)$ are error free, the
expected error in the color can be expressed as
\begin{equation}
\sigma(C) = \frac{2.5 \; \log_{10} {\rm e}}%
{\displaystyle\frac{\langle f(\lambda) \rangle_1}{\langle f(\lambda) \rangle_2}}
\;\; \sigma\left(
\displaystyle\frac{\langle f(\lambda) \rangle_1}{\langle f(\lambda)\rangle_2}
\right).
\label{equation_sigma_color}
\end{equation}
The square of the last term in the previous expression can
be written as
\begin{equation}
\begin{array}{@{}l}
\sigma^2\left(
\displaystyle\frac{\langle f(\lambda) \rangle_1}{\langle f(\lambda)\rangle_2}
\right) = \\ \noalign{\medskip} \;\;\;\;\;\;\;\;\; =
\displaystyle
\frac{\sigma^2(\langle f(\lambda) \rangle_1)}{(\langle f(\lambda) \rangle_2)^2}
\; + \;
\displaystyle
\frac{(\langle f(\lambda) \rangle_1)^2}{(\langle f(\lambda) \rangle_2)^2} 
\;
\displaystyle
\frac{\sigma^2(\langle f(\lambda) \rangle_2)}{(\langle f(\lambda)\rangle_2)^2}.
\end{array}
\label{equation_sigma_meanfluxes}
\end{equation}
In the other hand, from Eq.~(\ref{equation_color_meanfluxes}) it is immediate 
to show that
\begin{equation}
\displaystyle
\frac{(\langle f(\lambda) \rangle_1)^2}{(\langle f(\lambda) \rangle_2)^2} 
=
\left[10^{-0.4 (C-C_0)}\right]^2.
\label{equation_simple_fraction}
\end{equation}
The first fraction in the right hand side of
Eq.~(\ref{equation_sigma_meanfluxes}) can also be rewritten as
\begin{equation}
\begin{array}{@{}ll}
\displaystyle
\frac{\sigma^2(\langle f(\lambda) \rangle_1)}{(\langle f(\lambda) \rangle_2)^2}
& =
\displaystyle
\frac{(\langle f(\lambda) \rangle_1)^2}{(\langle f(\lambda) \rangle_2)^2} 
\;
\displaystyle
\frac{\sigma^2(\langle f(\lambda) \rangle_1)}{(\langle f(\lambda)\rangle_1)^2}
= \\ \noalign{\medskip}
& =
\left[10^{-0.4 (C-C_0)}\right]^2
\displaystyle
\frac{\sigma^2(\langle f(\lambda) \rangle_1)}{(\langle f(\lambda)\rangle_1)^2}.
\end{array}
\end{equation}
It is not difficult to show that assuming that the error spectrum is roughly
constant within each bandpass filter, making use of
Eqs.~(\ref{equation_meanflux}) and~(\ref{equation_ki}), and after replacing the
integrals by summations,
\begin{equation}
\displaystyle
\frac{\sigma^2(\langle f(\lambda) \rangle_i)}{(\langle f(\lambda)\rangle_i)^2}
\simeq
\left( \frac{\langle\sigma_{i,\mbox{\AA}}\rangle}%
{\langle f(\lambda)\rangle_{i,\mbox{\AA}}} \right)^2 \;
\xi_i \simeq \frac{1}{[{\it SN}(\mbox{\AA})_i]^2} \; \xi_i,
\label{equation_complex_fraction}
\end{equation}
where $\langle f(\lambda)\rangle_{i,\mbox{\AA}}$ and 
$\langle\sigma_{i,\mbox{\AA}}\rangle$ are the mean flux and mean error per \AA\ 
in the $i^{\rm th}$ filter, respectively, ${{\it SN}(\mbox{\AA})_i}$ the mean
signal-to-noise ratio per \AA\ in the $i^{\rm th}$ filter, and
\begin{equation}
\xi_i \equiv \frac{1}{\Theta} \;\;\;
\displaystyle\frac{\displaystyle\sum_{k=1}^{N_i} [ R_i(\lambda_{i,k})]^2}%
{\left[\displaystyle\sum_{k=1}^{N_i} R_i(\lambda_{i,k})\right]^2}.
\label{equation_xi_definition}
\end{equation}
Introducing the result of
Eqs.~(\ref{equation_sigma_meanfluxes})--(\ref{equation_complex_fraction}) into
Eq.~(\ref{equation_sigma_color}), we obtain
\begin{equation}
\begin{array}{@{}ll}
\sigma(C) = & 2.5 \; \log_{10} {\rm e} 
\\ \noalign{\medskip}
 & \times \; \left(
\displaystyle\frac{\xi_1}{[{\it SN}(\mbox{\AA})_1]^2} 
+ 
\displaystyle\frac{\xi_2}{[{\it SN}(\mbox{\AA})_2]^2} 
\right)^{1/2}.
\end{array}
\end{equation}
If ${{\it SN}(\mbox{\AA})_1} \approx {{\it SN}(\mbox{\AA})_2}$,
this last expression adopts the same form that
Eq.~(\ref{equation_error_sn}), 
\begin{equation}
\sigma(C) \simeq \frac{c(C)}{\mbox{{\it SN}({\mbox{\AA}})}},
\end{equation}
where
\begin{equation}
\begin{array}{@{}ll}
c(C) & = 2.5 \; \log_{10} {\rm e} \; \times \; (\xi_1 + \xi_2)^{1/2}
\\ \noalign{\medskip}
     & \simeq 1.086 \; (\xi_1 + \xi_2)^{1/2}.
\end{array}
\label{equation_xi1xi2}
\end{equation}
Numerical values of $\xi_i$ for typical photometric bands are given in
Table~\ref{table_xicolors}. 

\begin{figure}
 \resizebox{1.0\hsize}{!}{\includegraphics[angle=-90]{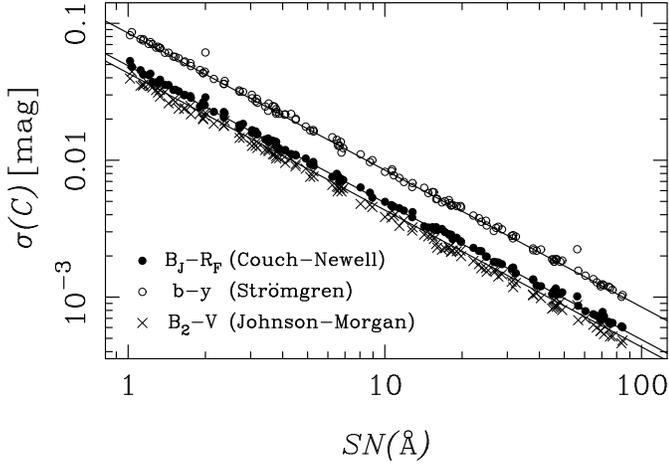}}
 \caption{Random errors from numerical simulations in the measurement of three
 colors in the 131 stellar spectra from the library of
 Pickles~(\cite{pickles98}), as a function of the signal-to-noise ratio per
 \AA. The full lines are the predictions of Eq.~(\ref{equation_xi1xi2}). 
 See text for details.}
 \label{plot_errsn}
\end{figure}

For illustration, we compare in Fig.~\ref{plot_errsn} the predictions of
Eq.~(\ref{equation_xi1xi2}), for three sample colors, with the results of
numerical simulations.  For this purpose, we have employed the 131 stellar
spectra from the library of Pickles~(\cite{pickles98}), which contains SEDs
with an ample range of spectral types and luminosity classes. For each of
these spectra, we have built a synthetic error spectrum by randomly choosing a
given $\mbox{{\it SN}({\mbox{\AA}})}$. Color errors were then measured in
simulated instances of the spectra generated with
Eq.~(\ref{equation_randomspectrum}). The full lines in Fig.~\ref{plot_errsn}
are not fits to the data points, but the predictions of
Eq.~(\ref{equation_xi1xi2}) using the corresponding $\xi_i$ parameters
(extracted from Table~\ref{table_xicolors}).

\begin{figure}
 \resizebox{1.0\hsize}{!}{\includegraphics{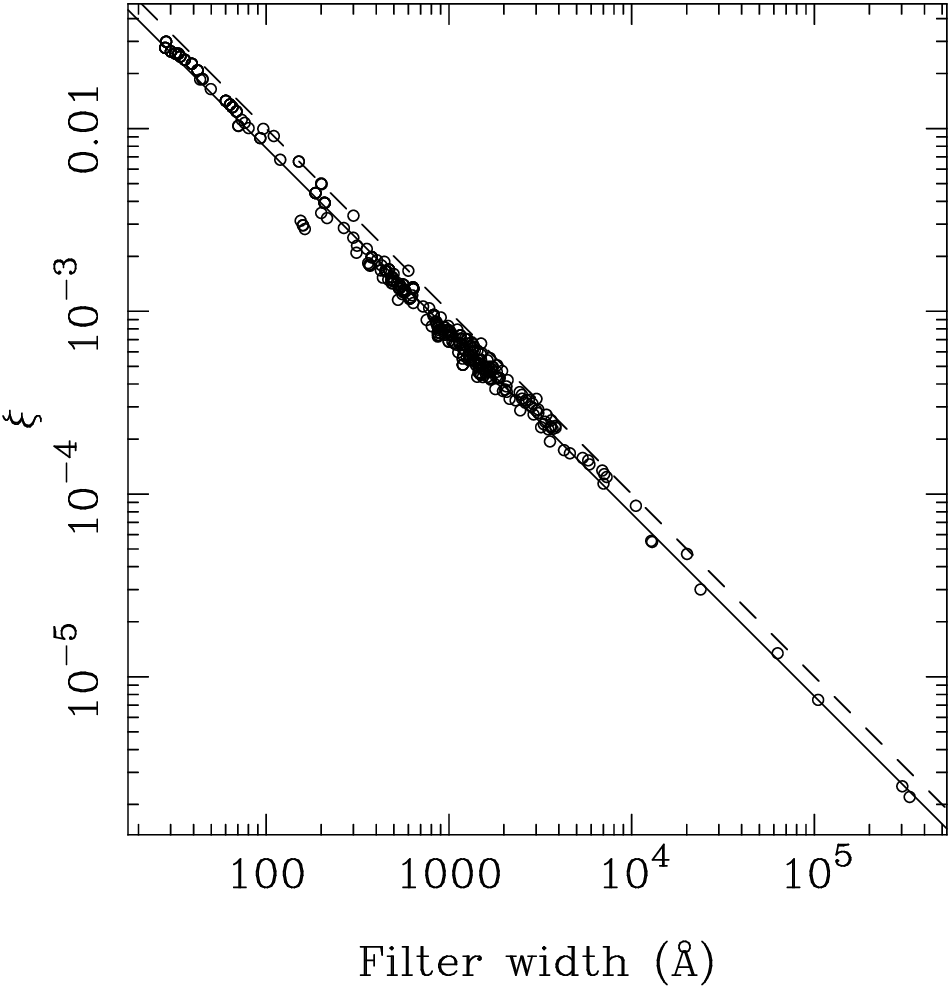}}
 \caption{Comparison of the $\xi$ values as a function of the filter width,
 computed with Eqs.~(\ref{equation_xi_definition})
 and~(\ref{equation_filterwidth}), respectively. The measurements have been
 performed using the response function of a set of 360 photometric bands
 collected from the literature (including those which are
 incorporated in the source code of the GISSEL models, and HST, KPNO and ESO
 filters, among others). The solid line is Eq.~(\ref{equation_xireal}), a least
 square fit to the data, and the dashed line is the prediction of
 Eq.~(\ref{equation_xiwidth}). The difference between both lines is due to the
 fact that filter response curves are not exactly box functions.}
 \label{diagrama_fwxi}
\end{figure}

Finally, it is also possible to express the $\xi_i$ coefficients as a function
of the filter width. In fact Eq.~(\ref{equation_xi_definition}) is the
discrete expression of the more general definition
\begin{equation}
\xi_i \equiv 
  \frac{\displaystyle\int_{{\rm filter}_i} R(\lambda)^2\;{\rm d}\lambda}%
{\left[\displaystyle\int_{{\rm filter}_i} R(\lambda)\;{\rm d}\lambda\right]^2}.
\label{equation_xigeneral}
\end{equation}
If now we assume that the filter transmission can be approximately described by
a box function of the form
\begin{equation}
R(\lambda) = \left\{
\begin{array}{@{}ll}
R_0 \;\;\mbox{(constant)} & \mbox{for $\lambda_1 \le \lambda \le \lambda_2$,} \\
0                         & \mbox{for $\lambda < \lambda_1$ 
                                  or $\lambda > \lambda_2$,}
\end{array}
\right.
\end{equation}
it is immediate to show that
\begin{equation}
\xi = (\lambda_2 - \lambda_1)^{-1} \equiv \mbox{\it FW}({\mbox{\AA}})^{-1},
\label{equation_xiwidth}
\end{equation}
where $\mbox{\it FW}({\mbox{\AA}})$ is the filter width in \AA.  In practice,
since filter response functions are not exactly box functions, filter widths
can be approximately computed as
\begin{equation}
\mbox{\it FW}({\mbox{\AA}}) \simeq \displaystyle\int_{{\rm filter}_i} 
                                   \tilde{R}(\lambda) \;
                                   {\rm d}\lambda,
\label{equation_filterwidth}
\end{equation}
being $\tilde{R}(\lambda)$ the response function normalized to unity.
To check the validity of these approximations, in Fig.~\ref{diagrama_fwxi} we
graphically compare the measured filter widths versus the $\xi$ values derived
using Eq.~(\ref{equation_xi_definition}), for a collection of 360 filter
response functions collected from the literature (see table caption). The full
line is a least squares fit, whereas the dashed line is the prediction of
Eq.~(\ref{equation_xiwidth}). As it is apparent from the figure, there is an
excellent correlation between the $\xi$ coefficients and the filter width,
although the linear fit to the data points (full line) indicates that, on
average,
\begin{equation}
\xi \simeq (0.78 \pm 0.09) \; \mbox{\it FW}({\mbox{\AA}})^{-1},
\label{equation_xireal}
\end{equation}
where the error in the coefficient is the residual standard deviation around
the fit.

\begin{table*}[t]
\begin{tabular}{lcrrclcrr} \hline \hline
bandpass system      & band  & \multicolumn{1}{c}{width} &
\multicolumn{1}{c}{$\xi$}  & \hspace{4mm} &
bandpass system      & band  & \multicolumn{1}{c}{width} &
\multicolumn{1}{c}{$\xi$}  \\
                     &       & \multicolumn{1}{c}{(\AA)} & 
\multicolumn{1}{c}{$\times 10^{-4}$} & &
                     &       & \multicolumn{1}{c}{(\AA)} & 
\multicolumn{1}{c}{$\times 10^{-4}$} \\ \hline
Johnson-Morgan       & $U_3$           &  543
 &  13.65 &  & 
Schneider et al.     & $g_4$           &  943
 &   8.17 \\
                     & $B_2$           & 1010
 &   7.70 &  & 
(4-shooter)          & $r_4$           &  906
 &   7.97 \\
                     & $V$             &  871
 &   8.15 &  & 
                     & $i_4$           & 1405
 &   5.06 \\
          &          &                 &        &  & 
                     & $z_4$           & 1286
 &   5.41 \\
Cousins              & $R_{\rm C}$     & 1591
 &   4.56 &  & 
          &          &                 &        \\
                     & $I_{\rm C}$     & 1495
 &   5.82 &  & 
Schneider et al.     & $g$             &  894
 &   7.60 \\
          &          &                 &        &  & 
(Pfuei)              & $r$             &  880
 &   7.90 \\
Johnson              & $R_{\rm J}$     & 1978
 &   3.66 &  & 
                     & $i$             & 1196
 &   5.73 \\
                     & $I_{\rm J}$     & 2148
 &   3.32 &  & 
                     & $z$             & 1187
 &   5.50 \\
          &          &                 &        &  & 
          &          &                 &        \\
Sandage-Smith        & $u$             &  580
 &  12.85 &  & 
Schneider et al.     & $A$             &  498
 &  16.01 \\
                     & $b$             & 1046
 &   7.34 &  & 
(narrow bands)       & $B$             &  443
 &  18.68 \\
                     & $v$             &  854
 &   8.76 &  & 
                     & $C$             &  571
 &  13.17 \\
                     & $r$             &  998
 &   6.84 &  & 
                     & $D$             &  491
 &  15.40 \\
          &          &                 &        &  & 
          &          &                 &        \\
Str\"{o}mgren        & $u$             &  355
 &  21.95 &  & 
Tyson (CCD)          & $B_{\rm J}$     & 1214
 &   6.16 \\
                     & $v$             &  200
 &  34.49 &  & 
                     & $R$             & 1307
 &   6.42 \\
                     & $b$             &  215
 &  32.36 &  & 
                     & $I$             & 1688
 &   4.30 \\
                     & $y$             &  266
 &  28.57 &  & 
          &          &                 &        \\
          &          &                 &        &  & 
WFPC2                & F555W           & 1489
 &   5.13 \\
Kron                 & $U_{\rm K}$     &  565
 &  12.72 &  & 
                     & F606W           & 1849
 &   4.32 \\
                     & $J_{\rm K}$     & 1297
 &   5.74 &  & 
                     & F702W           & 1662
 &   4.72 \\
                     & $F_{\rm K}$     & 1190
 &   5.09 &  & 
                     & F814W           & 1485
 &   4.52 \\
                     & $N_{\rm K}$     & 1646
 &   4.98 &  & 
          &          &                 &        \\
          &          &                 &        &  & 
POSS II              & $g_{\rm POSS}$  &  780
 &  10.40 \\
Couch-Newell         & $B_{\rm J}$     & 1384
 &   6.10 &  & 
                     & $r_{\rm POSS}$  &  991
 &   7.47 \\
                     & $R_{\rm F}$     &  486
 &  14.23 &  & 
                     & $i_{\rm POSS}$  & 1191
 &   6.25 \\
          &          &                 &        &  & 
          &          &                 &        \\
Thuan-Gunn           & $u$             &  404
 &  19.02 &  & 
SDSS                 & $u\prime$       &  568
 &  13.92 \\
                     & $v$             &  485
 &  15.33 &  & 
                     & $g\prime$       & 1264
 &   6.66 \\
                     & $g$             &  722
 &  10.63 &  & 
                     & $r\prime$       & 1333
 &   6.76 \\
                     & $r$             &  901
 &   9.27 &  & 
                     & $i\prime$       & 1349
 &   6.08 \\
          &          &                 &        &  & 
                     & $z\prime$       & 1309
 &   5.39 \\
\hline
\end{tabular}
\caption{Numerical values of the $\xi$ coefficients computed from
Eq.~(\ref{equation_xi_definition}), for a set of common photometric bands
corresponding to the filters given in Table~9 of Fukugita et
al.~(\cite{fukugita95}). Filter widths were determined with
Eq.~(\ref{equation_filterwidth}).}
\label{table_xicolors}
\end{table*}

\section{Measuring age and metallicity from index-index diagrams}
\label{appendix_sensitivity}

Iso-metallicity and iso-age lines in index-index, index-color and color-color
diagrams are usually far from displaying very regular grids, but they typically
exhibit unevenly spaced and distorted patterns. For this reason, the
computation of ages and metallicities from a given pair of spectroscopic
measurements should be addressed through the use of local mapping functions
which properly accounts for the geometric distortions. Obviously, the same is
also true for the computation of the local derivatives (i.e.\ metal sensitivity
parameters). 

An excellent approach to this problem is the use of bivariate polynomial
transformations of the form (see e.g.\ Wolberg 1992)
\begin{equation}
\begin{array}{@{}r@{\;}c@{\;}l}
  m_1 & = & \displaystyle\sum_{i=0}^{N} \sum_{j=0}^{N-i} \; p_{ij} \; 
    (\log[{\rm age}])^i \; (\log[Z])^j, 
\\ \noalign{\medskip}
  m_2 & = & \displaystyle\sum_{i=0}^{N} \sum_{j=0}^{N-i} \; q_{ij} \; 
    (\log[{\rm age}])^i \; (\log[Z])^j,
\end{array}
\label{eq_bivariate}
\end{equation}
where $N$ is the polynomial degree, and $p_{ij}$ and $q_{ij}$ are the
polynomial coefficients. The inverse transformation can be written by an
analogous expression.

\begin{figure}
 \resizebox{1.0\hsize}{!}{%
 \includegraphics[bb=108 138 571 594,angle=-90]{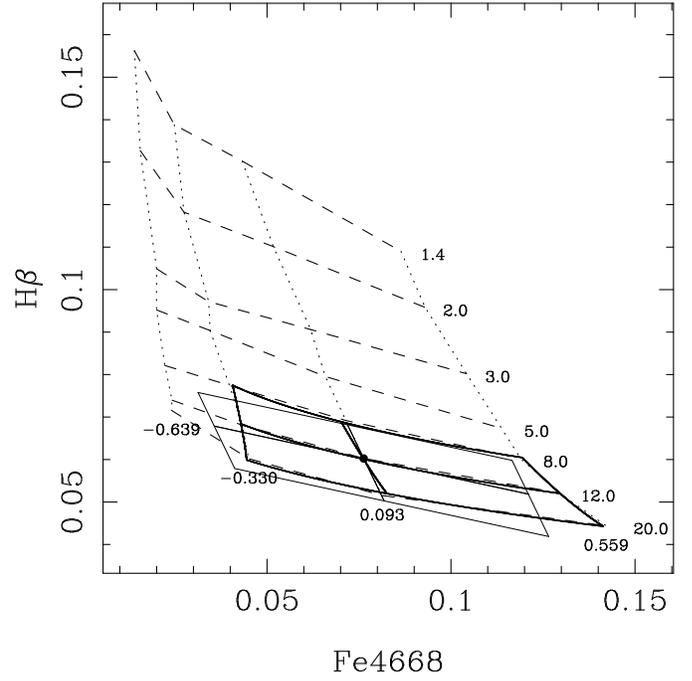}}
 \caption{Example of fit of local polynomials to describe the local behavior of
 line-strength indices as a function of physical parameters.  Dashed and dotted
 lines are the predictions of Bruzual \& Charlot (2001) models for single
 stellar populations of fixed age and metallicity, respectively (ages are given
 in Gyr, and metallicities as [Fe/H]). The thin solid lines indicates the
 resulting fit after using Eq.~(\ref{eq_bivariate}) with $N=1$, around the
 model predictions for a SSP of 12~Gyr and solar metallicity (the
 coefficients were determined using a least-squares fit to 5 points: the grid
 model point chosen as the origin of the local transformation, and the two
 closest points in both age and metallicity).  It is clear that the fit can
 not be extrapolated very far from the central point. The thick solid lines
 show the result for $N=2$ (derived from the fit to the 9 points: the 5 points
 previously employed in the fit for $N=1$, plus the 4 additional corners of the
 region delineated by the thick line). In this case it is clear that the $N=2$
 polynomial approximation provides a very reasonable representation of the
 geometric distortions when moving from the observational to the physical
 parameter space.}
 \label{diagrama_Hbeta_Fe4668}
\end{figure}

For $N=1$ the above equations only account for affine transformations (i.e.\
translation, rotation, scale and shear).  This linear approximation is valid
when the 6 polynomial coefficients are derived from control points (those for
which line-strength indices, ages and metallicities are given by the models)
which are very close to the point ($p_{00},q_{00}$). Although most evolutionary
synthesis models provide these close control points when considering the
line-strength predictions as a function of age, the same is not true for the
indices variations as a function of metallicity. This problem leads to
systematic uncertainties in the index predictions, as shown in
Fig.~\ref{diagrama_Hbeta_Fe4668} for the H$\beta$-Fe4668 diagram.

The second-degree approximation, $N=2$, improves the quality of the prediction
allowing for a selection of more distant control points. In this case, 12
coefficients must be computed by solving two systems of 6 linear equations. It
is straightforward to show that the coefficients of the $A$ matrix in
Eq.~(\ref{eq_linear_approx}) can be rewritten as a function of the bivariate
polynomial coefficients as
\begin{equation}
\begin{array}{@{}r@{\;}c@{\;}l} 
a_{11} & = & p_{10}+p_{11}\log[{\rm age}]+2 \; p_{20}\log[Z], \\
a_{12} & = & p_{01}+p_{11}\log[Z]+2 \; p_{02}\log[{\rm age}], \\
a_{21} & = & q_{10}+q_{11}\log[{\rm age}]+2 \; q_{20}\log[Z], \\
a_{22} & = & q_{01}+q_{11}\log[Z]+2 \; q_{02}\log[{\rm age}]. \\
\end{array}
\label{matrixA_bivariate}
\end{equation}
In Fig.~\ref{diagrama_Hbeta_Fe4668} we also show the mapping obtained for $N=2$
(thick solid lines), which clearly improves the result obtained for $N=1$.
Higher order polynomials ($N=3,...$) are typically unnecessary, since the
required number of points to perform the fit increases rapidly, and in these
situations it is always possible to constraint the fit to a smaller region with
$N=2$.

Obviously, the same procedure is valid when reading other physical
parameters from other grids predicted by stellar population models.

%
\end{document}

%% file: rmodel1.inc
\begin{tabular}{c|@{$\;$}r@{$\;$}r@{$\;$}r@{$\;$}r@{$\;$}r@{$\;$}r@{$\;$}r@{$\;$}r@{$\;$}r@{$\;$}r@{$\;$}r@{$\;$}r@{$\;$}r@{$\;$}r@{$\;$}r@{$\;$}r@{$\;$}r@{$\;$}r@{$\;$}r@{$\;$}r@{$\;$}r@{$\;$}r} \hline \hline
 & \multicolumn{1}{c}{\rotatebox{90}{$\tilde{\rm D}_{4000}$}} & \multicolumn{1}{c}{\rotatebox{90}{CN$_1$}} & \multicolumn{1}{c}{\rotatebox{90}{CN$_2$}} & \multicolumn{1}{c}{\rotatebox{90}{Ca4227}} & \multicolumn{1}{c}{\rotatebox{90}{G4300}} & \multicolumn{1}{c}{\rotatebox{90}{Fe4383}} & \multicolumn{1}{c}{\rotatebox{90}{Ca4455}} & \multicolumn{1}{c}{\rotatebox{90}{Fe4531}} & \multicolumn{1}{c}{\rotatebox{90}{Fe4668}} & \multicolumn{1}{c}{\rotatebox{90}{H$\beta$}} & \multicolumn{1}{c}{\rotatebox{90}{Fe5015}} & \multicolumn{1}{c}{\rotatebox{90}{Mg$_1$}} & \multicolumn{1}{c}{\rotatebox{90}{Mg$_2$}} & \multicolumn{1}{c}{\rotatebox{90}{Mgb}} & \multicolumn{1}{c}{\rotatebox{90}{Fe5270}} & \multicolumn{1}{c}{\rotatebox{90}{Fe5335}} & \multicolumn{1}{c}{\rotatebox{90}{Fe5406}} & \multicolumn{1}{c}{\rotatebox{90}{Fe5709}} & \multicolumn{1}{c}{\rotatebox{90}{Fe5782}} & \multicolumn{1}{c}{\rotatebox{90}{Na5895}} & \multicolumn{1}{c}{\rotatebox{90}{TiO$_1$}} & \multicolumn{1}{c}{\rotatebox{90}{TiO$_2$}}\\ \hline
$\tilde{\rm D}_{4000}$ & \ldots & \underline{\boldmath$0.2$} & \underline{\boldmath$0.4$} & $1.2$ & $1.9$ & \underline{\boldmath$0.3$} & \underline{\boldmath$0.6$} & \underline{\boldmath$0.8$} & \underline{\boldmath$0.0$} & \underline{\boldmath$0.5$} & \underline{\boldmath$0.4$} & \underline{\boldmath$0.2$} & \underline{\boldmath$0.0$} & \underline{\boldmath$0.5$} & \underline{\boldmath$0.6$} & \underline{\boldmath$0.4$} & \underline{\boldmath$0.5$} & \underline{\boldmath$0.6$} & \underline{\boldmath$0.7$} & \underline{\boldmath$0.3$} & $1.4$ & \underline{\boldmath$0.9$}\\ 
CN$_1$ & \ldots & \ldots & $1.7$ & $1.5$ & $1.5$ & $1.5$ & $1.8$ & $2.4$ & \underline{\boldmath$0.9$} & $1.2$ & $1.4$ & $2.8$ & $1.9$ & $2.5$ & $1.9$ & $1.5$ & $1.6$ & $1.5$ & $1.6$ & $1.6$ & $1.7$ & $1.6$\\ 
CN$_2$ & \ldots & \ldots & \ldots & $1.7$ & $1.6$ & $1.4$ & $1.7$ & $2.1$ & \underline{\boldmath$0.9$} & $1.2$ & $1.4$ & $1.7$ & $1.4$ & $1.9$ & $1.7$ & $1.4$ & $1.5$ & $1.5$ & $1.6$ & $1.5$ & $1.7$ & $1.8$\\ 
Ca4227 & \ldots & \ldots & \ldots & \ldots & $2.4$ & $1.5$ & $1.8$ & $2.0$ & $1.2$ & $1.6$ & $1.6$ & $1.5$ & $1.3$ & $1.8$ & $1.8$ & $1.6$ & $1.7$ & $1.7$ & $1.8$ & $1.5$ & $2.3$ & $2.3$\\ 
G4300 & \ldots & \ldots & \ldots & \ldots & \ldots & $1.6$ & $1.9$ & $2.1$ & $1.3$ & $1.9$ & $1.7$ & $1.5$ & $1.3$ & $1.8$ & $1.9$ & $1.7$ & $1.8$ & $1.9$ & $2.0$ & $1.6$ & $2.8$ & $2.1$\\ 
Fe4383 & \ldots & \ldots & \ldots & \ldots & \ldots & \ldots & $3.3$ & $2.3$ & $1.4$ & $1.4$ & $1.9$ & $1.5$ & $1.4$ & $1.8$ & $2.8$ & $2.0$ & $2.3$ & $1.9$ & $2.1$ & $2.1$ & $1.8$ & $1.5$\\ 
Ca4455 & \ldots & \ldots & \ldots & \ldots & \ldots & \ldots & \ldots & $2.6$ & $1.7$ & $1.7$ & $2.2$ & $1.8$ & $1.7$ & $2.2$ & $3.3$ & $2.4$ & $2.6$ & $2.2$ & $2.4$ & $2.5$ & $2.1$ & $1.9$\\ 
Fe4531 & \ldots & \ldots & \ldots & \ldots & \ldots & \ldots & \ldots & \ldots & $1.6$ & $1.8$ & $2.1$ & $2.3$ & $2.5$ & $2.9$ & $2.7$ & $2.2$ & $2.3$ & $2.2$ & $2.3$ & $2.6$ & $2.2$ & $2.1$\\ 
Fe4668 & \ldots & \ldots & \ldots & \ldots & \ldots & \ldots & \ldots & \ldots & \ldots & $1.2$ & $2.4$ & \underline{\boldmath$0.9$} & \underline{\boldmath$0.8$} & $1.3$ & $1.6$ & $1.9$ & $1.8$ & $2.3$ & $3.1$ & $1.2$ & $1.5$ & $1.2$\\ 
H$\beta$ & \ldots & \ldots & \ldots & \ldots & \ldots & \ldots & \ldots & \ldots & \ldots & \ldots & $1.7$ & $1.2$ & \underline{\boldmath$1.0$} & $1.5$ & $1.7$ & $1.6$ & $1.7$ & $1.8$ & $1.9$ & $1.3$ & $2.2$ & $1.6$\\ 
Fe5015 & \ldots & \ldots & \ldots & \ldots & \ldots & \ldots & \ldots & \ldots & \ldots & \ldots & \ldots & $1.4$ & $1.3$ & $1.7$ & $2.1$ & $2.5$ & $2.4$ & $2.5$ & $3.4$ & $1.7$ & $1.9$ & $1.6$\\ 
Mg$_1$ & \ldots & \ldots & \ldots & \ldots & \ldots & \ldots & \ldots & \ldots & \ldots & \ldots & \ldots & \ldots & $1.8$ & $2.4$ & $1.8$ & $1.5$ & $1.6$ & $1.5$ & $1.6$ & $1.6$ & $1.6$ & $1.6$\\ 
Mg$_2$ & \ldots & \ldots & \ldots & \ldots & \ldots & \ldots & \ldots & \ldots & \ldots & \ldots & \ldots & \ldots & \ldots & $2.8$ & $1.8$ & $1.3$ & $1.5$ & $1.3$ & $1.5$ & $1.6$ & $1.5$ & $1.4$\\ 
Mgb & \ldots & \ldots & \ldots & \ldots & \ldots & \ldots & \ldots & \ldots & \ldots & \ldots & \ldots & \ldots & \ldots & \ldots & $2.3$ & $1.8$ & $1.9$ & $1.8$ & $2.0$ & $2.0$ & $2.0$ & $1.9$\\ 
Fe5270 & \ldots & \ldots & \ldots & \ldots & \ldots & \ldots & \ldots & \ldots & \ldots & \ldots & \ldots & \ldots & \ldots & \ldots & \ldots & $2.3$ & $2.5$ & $2.1$ & $2.3$ & $2.7$ & $2.1$ & $1.9$\\ 
Fe5335 & \ldots & \ldots & \ldots & \ldots & \ldots & \ldots & \ldots & \ldots & \ldots & \ldots & \ldots & \ldots & \ldots & \ldots & \ldots & \ldots & $2.8$ & $2.3$ & $2.6$ & $1.8$ & $1.9$ & $1.6$\\ 
Fe5406 & \ldots & \ldots & \ldots & \ldots & \ldots & \ldots & \ldots & \ldots & \ldots & \ldots & \ldots & \ldots & \ldots & \ldots & \ldots & \ldots & \ldots & $2.3$ & $2.6$ & $2.0$ & $2.0$ & $1.7$\\ 
Fe5709 & \ldots & \ldots & \ldots & \ldots & \ldots & \ldots & \ldots & \ldots & \ldots & \ldots & \ldots & \ldots & \ldots & \ldots & \ldots & \ldots & \ldots & \ldots & $2.9$ & $1.7$ & $2.1$ & $1.7$\\ 
Fe5782 & \ldots & \ldots & \ldots & \ldots & \ldots & \ldots & \ldots & \ldots & \ldots & \ldots & \ldots & \ldots & \ldots & \ldots & \ldots & \ldots & \ldots & \ldots & \ldots & $1.9$ & $2.2$ & $1.8$\\ 
Na5895 & \ldots & \ldots & \ldots & \ldots & \ldots & \ldots & \ldots & \ldots & \ldots & \ldots & \ldots & \ldots & \ldots & \ldots & \ldots & \ldots & \ldots & \ldots & \ldots & \ldots & $1.8$ & $1.6$\\ 
TiO$_1$ & \ldots & \ldots & \ldots & \ldots & \ldots & \ldots & \ldots & \ldots & \ldots & \ldots & \ldots & \ldots & \ldots & \ldots & \ldots & \ldots & \ldots & \ldots & \ldots & \ldots & \ldots & $2.2$\\ 
TiO$_2$ & \ldots & \ldots & \ldots & \ldots & \ldots & \ldots & \ldots & \ldots & \ldots & \ldots & \ldots & \ldots & \ldots & \ldots & \ldots & \ldots & \ldots & \ldots & \ldots & \ldots & \ldots & \ldots\\ 
\hline
\end{tabular}

%% file: rmodel2.inc
\begin{tabular}{c|@{$\;$}r@{$\;$}r@{$\;$}r@{$\;$}r@{$\;$}r@{$\;$}r@{$\;$}r@{$\;$}r@{$\;$}r@{$\;$}r@{$\;$}r@{$\;$}r@{$\;$}r@{$\;$}r@{$\;$}r@{$\;$}r} \hline \hline
 & \multicolumn{1}{c}{\rotatebox{90}{$(u-g)_{\rm AB}$}} & \multicolumn{1}{c}{\rotatebox{90}{$(g-r)_{\rm AB}$}} & \multicolumn{1}{c}{\rotatebox{90}{$(g-i)_{\rm AB}$}} & \multicolumn{1}{c}{\rotatebox{90}{$(g-z)_{\rm AB}$}} & \multicolumn{1}{c}{\rotatebox{90}{$(g-J)_{\rm AB}$}} & \multicolumn{1}{c}{\rotatebox{90}{$(g-H)_{\rm AB}$}} & \multicolumn{1}{c}{\rotatebox{90}{$(g-Ks)_{\rm AB}$}} & \multicolumn{1}{c}{\rotatebox{90}{$(U-B)$}} & \multicolumn{1}{c}{\rotatebox{90}{$(B-V)$}} & \multicolumn{1}{c}{\rotatebox{90}{$(V-R)$}} & \multicolumn{1}{c}{\rotatebox{90}{$(V-I)$}} & \multicolumn{1}{c}{\rotatebox{90}{$(V-J)$}} & \multicolumn{1}{c}{\rotatebox{90}{$(V-K)$}} & \multicolumn{1}{c}{\rotatebox{90}{$(R-I)$}} & \multicolumn{1}{c}{\rotatebox{90}{$(J-H)$}} & \multicolumn{1}{c}{\rotatebox{90}{$(H-K)$}}\\ \hline
$(u-g)_{\rm AB}$ & \ldots & $-1.1$ & $-1.6$ & $-1.6$ & $-1.7$ & $-1.9$ & \underline{\boldmath$-2.0$} & $-0.8$ & $-1.1$ & $-1.1$ & $-1.6$ & $-1.8$ & \underline{\boldmath$-2.1$} & $-1.6$ & $-1.1$ & $-1.7$\\ 
$(g-r)_{\rm AB}$ & \ldots & \ldots & $-1.0$ & $-0.6$ & $-1.6$ & $-1.7$ & $-1.9$ & $-1.2$ & $0.7$ & $-0.4$ & $-1.1$ & $-1.6$ & $-1.9$ & $-1.1$ & $-0.9$ & $-1.4$\\ 
$(g-i)_{\rm AB}$ & \ldots & \ldots & \ldots & $-1.2$ & $-2.0$ & \underline{\boldmath$-2.1$} & \underline{\boldmath$-2.2$} & $-1.6$ & $-1.0$ & $-0.4$ & $-0.8$ & $-1.9$ & \underline{\boldmath$-2.1$} & $-1.1$ & $-1.1$ & $-1.6$\\ 
$(g-z)_{\rm AB}$ & \ldots & \ldots & \ldots & \ldots & \underline{\boldmath$-2.0$} & \underline{\boldmath$-2.1$} & \underline{\boldmath$-2.2$} & $-1.6$ & $-0.5$ & $-0.4$ & $-1.3$ & $-2.0$ & \underline{\boldmath$-2.2$} & $-1.4$ & $-1.2$ & $-1.8$\\ 
$(g-J)_{\rm AB}$ & \ldots & \ldots & \ldots & \ldots & \ldots & $-1.3$ & $-1.8$ & $-1.5$ & $-1.6$ & $-1.5$ & $-1.9$ & $-1.5$ & $-2.0$ & $-1.8$ & $-1.1$ & $-1.8$\\ 
$(g-H)_{\rm AB}$ & \ldots & \ldots & \ldots & \ldots & \ldots & \ldots & $-1.7$ & $-1.7$ & $-1.7$ & $-1.6$ & \underline{\boldmath$-2.0$} & $-1.4$ & $-2.0$ & $-1.9$ & $-1.1$ & $-1.9$\\ 
$(g-Ks)_{\rm AB}$ & \ldots & \ldots & \ldots & \ldots & \ldots & \ldots & \ldots & $-1.9$ & $-1.8$ & $-1.7$ & \underline{\boldmath$-2.1$} & $-0.9$ & $-1.7$ & $-2.0$ & $-1.0$ & $-1.9$\\ 
$(U-B)$ & \ldots & \ldots & \ldots & \ldots & \ldots & \ldots & \ldots & \ldots & $-1.1$ & $-1.1$ & $-1.6$ & $-1.6$ & $-1.9$ & $-1.5$ & $-1.0$ & $-1.6$\\ 
$(B-V)$ & \ldots & \ldots & \ldots & \ldots & \ldots & \ldots & \ldots & \ldots & \ldots & $-0.3$ & $-1.0$ & $-1.6$ & $-1.8$ & $-1.1$ & $-0.8$ & $-1.4$\\ 
$(V-R)$ & \ldots & \ldots & \ldots & \ldots & \ldots & \ldots & \ldots & \ldots & \ldots & \ldots & $-0.6$ & $-1.5$ & $-1.7$ & $-0.8$ & $-0.7$ & $-1.2$\\ 
$(V-I)$ & \ldots & \ldots & \ldots & \ldots & \ldots & \ldots & \ldots & \ldots & \ldots & \ldots & \ldots & $-1.9$ & \underline{\boldmath$-2.1$} & $-0.8$ & $-1.1$ & $-1.5$\\ 
$(V-J)$ & \ldots & \ldots & \ldots & \ldots & \ldots & \ldots & \ldots & \ldots & \ldots & \ldots & \ldots & \ldots & $-1.6$ & $-1.8$ & $-0.8$ & $-1.7$\\ 
$(V-K)$ & \ldots & \ldots & \ldots & \ldots & \ldots & \ldots & \ldots & \ldots & \ldots & \ldots & \ldots & \ldots & \ldots & $-1.9$ & $-0.7$ & $-1.7$\\ 
$(R-I)$ & \ldots & \ldots & \ldots & \ldots & \ldots & \ldots & \ldots & \ldots & \ldots & \ldots & \ldots & \ldots & \ldots & \ldots & $-0.9$ & $-1.3$\\ 
$(J-H)$ & \ldots & \ldots & \ldots & \ldots & \ldots & \ldots & \ldots & \ldots & \ldots & \ldots & \ldots & \ldots & \ldots & \ldots & \ldots & $-0.6$\\ 
$(H-K)$ & \ldots & \ldots & \ldots & \ldots & \ldots & \ldots & \ldots & \ldots & \ldots & \ldots & \ldots & \ldots & \ldots & \ldots & \ldots & \ldots\\ 
\hline
\end{tabular}

%% file: rmodel3.inc
\begin{tabular}{c|@{$\;$}r@{$\;$}r@{$\;$}r@{$\;$}r@{$\;$}r@{$\;$}r@{$\;$}r@{$\;$}r@{$\;$}r@{$\;$}r@{$\;$}r@{$\;$}r@{$\;$}r@{$\;$}r@{$\;$}r@{$\;$}r@{$\;$}r@{$\;$}r@{$\;$}r@{$\;$}r@{$\;$}r@{$\;$}r} \hline \hline
 & \multicolumn{1}{c}{\rotatebox{90}{$\tilde{\rm D}_{4000}$}} & \multicolumn{1}{c}{\rotatebox{90}{CN$_1$}} & \multicolumn{1}{c}{\rotatebox{90}{CN$_2$}} & \multicolumn{1}{c}{\rotatebox{90}{Ca4227}} & \multicolumn{1}{c}{\rotatebox{90}{G4300}} & \multicolumn{1}{c}{\rotatebox{90}{Fe4383}} & \multicolumn{1}{c}{\rotatebox{90}{Ca4455}} & \multicolumn{1}{c}{\rotatebox{90}{Fe4531}} & \multicolumn{1}{c}{\rotatebox{90}{Fe4668}} & \multicolumn{1}{c}{\rotatebox{90}{H$\beta$}} & \multicolumn{1}{c}{\rotatebox{90}{Fe5015}} & \multicolumn{1}{c}{\rotatebox{90}{Mg$_1$}} & \multicolumn{1}{c}{\rotatebox{90}{Mg$_2$}} & \multicolumn{1}{c}{\rotatebox{90}{Mgb}} & \multicolumn{1}{c}{\rotatebox{90}{Fe5270}} & \multicolumn{1}{c}{\rotatebox{90}{Fe5335}} & \multicolumn{1}{c}{\rotatebox{90}{Fe5406}} & \multicolumn{1}{c}{\rotatebox{90}{Fe5709}} & \multicolumn{1}{c}{\rotatebox{90}{Fe5782}} & \multicolumn{1}{c}{\rotatebox{90}{Na5895}} & \multicolumn{1}{c}{\rotatebox{90}{TiO$_1$}} & \multicolumn{1}{c}{\rotatebox{90}{TiO$_2$}}\\ \hline
$(u-g)_{\rm AB}$ & \underline{\boldmath$-0.7$} & $-0.2$ & $-0.0$ & $1.1$ & $0.5$ & $-0.2$ & $0.1$ & $0.3$ & \underline{\boldmath$-0.6$} & $-0.1$ & $-0.1$ & $-0.2$ & \underline{\boldmath$-0.4$} & $0.1$ & $0.1$ & $-0.1$ & $-0.0$ & $-0.0$ & $0.1$ & $-0.2$ & $0.5$ & $0.8$\\ 
$(g-r)_{\rm AB}$ & $-0.1$ & $-0.1$ & $-0.0$ & $0.7$ & $1.4$ & $-0.0$ & $0.3$ & $0.5$ & $-0.3$ & $0.3$ & $0.1$ & $-0.1$ & $-0.3$ & $0.2$ & $0.3$ & $0.1$ & $0.2$ & $0.3$ & $0.4$ & $-0.0$ & $1.4$ & $0.5$\\ 
$(g-i)_{\rm AB}$ & \underline{\boldmath$-0.9$} & $-0.4$ & $-0.4$ & $0.2$ & $0.5$ & $-0.3$ & $0.1$ & $0.2$ & \underline{\boldmath$-0.5$} & $0.3$ & $-0.1$ & $-0.4$ & \underline{\boldmath$-0.6$} & $-0.1$ & $0.0$ & $-0.1$ & $-0.0$ & $0.1$ & $0.2$ & $-0.3$ & $1.1$ & $0.0$\\ 
$(g-z)_{\rm AB}$ & \underline{\boldmath$-0.7$} & \underline{\boldmath$-0.5$} & $-0.4$ & $0.2$ & $0.8$ & $-0.4$ & $-0.0$ & $0.1$ & \underline{\boldmath$-0.7$} & $0.0$ & $-0.2$ & \underline{\boldmath$-0.5$} & \underline{\boldmath$-0.7$} & $-0.2$ & $-0.1$ & $-0.2$ & $-0.1$ & $-0.1$ & $0.0$ & $-0.4$ & $1.4$ & $0.1$\\ 
$(g-J)_{\rm AB}$ & \underline{\boldmath$-1.3$} & $0.4$ & $0.4$ & $-0.0$ & $-0.0$ & $-0.2$ & $0.2$ & $0.6$ & \underline{\boldmath$-0.7$} & $-0.4$ & $-0.2$ & $0.5$ & $-0.0$ & $0.5$ & $0.2$ & $-0.1$ & $-0.0$ & $-0.1$ & $0.0$ & $-0.0$ & $0.1$ & $0.1$\\ 
$(g-H)_{\rm AB}$ & \underline{\boldmath$-1.4$} & $1.1$ & $0.1$ & $-0.1$ & $-0.1$ & $-0.1$ & $0.2$ & $0.8$ & \underline{\boldmath$-0.7$} & \underline{\boldmath$-0.4$} & $-0.2$ & $0.9$ & $0.3$ & $1.0$ & $0.2$ & $-0.2$ & $-0.0$ & $-0.2$ & $-0.0$ & $0.0$ & $0.0$ & $-0.0$\\ 
$(g-Ks)_{\rm AB}$ & \underline{\boldmath$-1.5$} & $-0.1$ & $-0.3$ & $-0.3$ & $-0.3$ & $0.1$ & $0.4$ & $1.0$ & \underline{\boldmath$-0.7$} & \underline{\boldmath$-0.5$} & $-0.2$ & $-0.1$ & $-0.1$ & $0.3$ & $0.5$ & $-0.1$ & $0.1$ & $-0.2$ & $0.0$ & $0.5$ & $-0.1$ & $-0.3$\\ 
$(U-B)$ & \underline{\boldmath$-0.7$} & $-0.0$ & $0.2$ & $0.7$ & $0.5$ & $-0.1$ & $0.2$ & $0.5$ & \underline{\boldmath$-0.5$} & $-0.1$ & $-0.0$ & $-0.0$ & $-0.3$ & $0.2$ & $0.2$ & $-0.0$ & $0.1$ & $0.1$ & $0.2$ & $-0.1$ & $0.5$ & $2.4$\\ 
$(B-V)$ & $-0.1$ & $-0.1$ & $0.0$ & $0.7$ & $1.4$ & $-0.0$ & $0.3$ & $0.5$ & $-0.3$ & $0.3$ & $0.2$ & $-0.1$ & $-0.3$ & $0.2$ & $0.3$ & $0.2$ & $0.2$ & $0.3$ & $0.4$ & $-0.0$ & $1.4$ & $0.5$\\ 
$(V-R)$ & $-0.3$ & $0.0$ & $0.1$ & $0.7$ & $1.1$ & $0.2$ & $0.5$ & $0.6$ & $-0.1$ & $0.6$ & $0.3$ & $0.0$ & $-0.2$ & $0.3$ & $0.5$ & $0.3$ & $0.4$ & $0.5$ & $0.6$ & $0.1$ & $2.1$ & $0.5$\\ 
$(V-I)$ & \underline{\boldmath$-0.9$} & $-0.4$ & $-0.3$ & $0.2$ & $0.4$ & $-0.2$ & $0.1$ & $0.2$ & \underline{\boldmath$-0.4$} & $0.5$ & $-0.0$ & $-0.4$ & \underline{\boldmath$-0.6$} & $-0.1$ & $0.1$ & $-0.0$ & $0.0$ & $0.2$ & $0.2$ & $-0.3$ & $0.9$ & $0.1$\\ 
$(V-J)$ & \underline{\boldmath$-1.3$} & $0.2$ & $-0.1$ & $-0.1$ & $-0.0$ & $0.2$ & $0.6$ & $1.7$ & \underline{\boldmath$-0.5$} & $-0.3$ & $0.0$ & $0.2$ & $0.3$ & $0.7$ & $0.7$ & $0.1$ & $0.2$ & $0.1$ & $0.2$ & $0.5$ & $0.1$ & $-0.0$\\ 
$(V-K)$ & \underline{\boldmath$-1.6$} & $-0.3$ & \underline{\boldmath$-0.5$} & $-0.3$ & $-0.3$ & $1.2$ & $3.4$ & $0.5$ & \underline{\boldmath$-0.5$} & \underline{\boldmath$-0.5$} & $0.0$ & $-0.4$ & \underline{\boldmath$-0.4$} & $0.0$ & $1.1$ & $0.2$ & $0.4$ & $-0.0$ & $0.2$ & $0.3$ & $-0.1$ & $-0.3$\\ 
$(R-I)$ & \underline{\boldmath$-0.9$} & $-0.3$ & $-0.2$ & $0.2$ & $0.4$ & $-0.1$ & $0.2$ & $0.3$ & $-0.3$ & $1.1$ & $0.1$ & $-0.3$ & \underline{\boldmath$-0.5$} & $0.0$ & $0.2$ & $0.1$ & $0.2$ & $0.3$ & $0.4$ & $-0.1$ & $0.8$ & $0.1$\\ 
$(J-H)$ & \underline{\boldmath$-0.6$} & $0.5$ & $0.4$ & $0.6$ & $0.7$ & $1.2$ & $1.5$ & $1.3$ & $0.7$ & $0.6$ & $1.3$ & $0.5$ & $0.4$ & $0.9$ & $1.4$ & $1.7$ & $2.5$ & $1.2$ & $1.4$ & $0.9$ & $0.9$ & $0.6$\\ 
$(H-K)$ & \underline{\boldmath$-1.1$} & $-0.2$ & $-0.2$ & $0.0$ & $0.2$ & $0.1$ & $0.4$ & $0.4$ & $0.3$ & $0.2$ & $0.7$ & $-0.3$ & $-0.4$ & $0.1$ & $0.4$ & $0.5$ & $0.5$ & $1.3$ & $1.0$ & $0.0$ & $0.4$ & $0.0$\\ 
\hline
\end{tabular}

%% file: rmodel_comparison.inc_modified
\begin{tabular}{cc|@{$\;$}r@{$\;$}r@{$\;$}r@{$\;$}r@{$\;$}r@{$\;$}r@{$\;$}r@{$\;$}r@{$\;$}r@{$\;$}r@{$\;$}r@{$\;$}r@{$\;$}r@{$\;$}r@{$\;$}r@{$\;$}r@{$\;$}r@{$\;$}r@{$\;$}r@{$\;$}r@{$\;$}r@{$\;$}r} \hline \hline
 index & models  & \multicolumn{1}{c}{\rotatebox{90}{$\tilde{\rm D}_{4000}$}} & \multicolumn{1}{c}{\rotatebox{90}{CN$_1$}} & \multicolumn{1}{c}{\rotatebox{90}{CN$_2$}} & \multicolumn{1}{c}{\rotatebox{90}{Ca4227}} & \multicolumn{1}{c}{\rotatebox{90}{G4300}} & \multicolumn{1}{c}{\rotatebox{90}{Fe4383}} & \multicolumn{1}{c}{\rotatebox{90}{Ca4455}} & \multicolumn{1}{c}{\rotatebox{90}{Fe4531}} & \multicolumn{1}{c}{\rotatebox{90}{Fe4668}} & \multicolumn{1}{c}{\rotatebox{90}{H$\beta$}} & \multicolumn{1}{c}{\rotatebox{90}{Fe5015}} & \multicolumn{1}{c}{\rotatebox{90}{Mg$_1$}} & \multicolumn{1}{c}{\rotatebox{90}{Mg$_2$}} & \multicolumn{1}{c}{\rotatebox{90}{Mgb}} & \multicolumn{1}{c}{\rotatebox{90}{Fe5270}} & \multicolumn{1}{c}{\rotatebox{90}{Fe5335}} & \multicolumn{1}{c}{\rotatebox{90}{Fe5406}} & \multicolumn{1}{c}{\rotatebox{90}{Fe5709}} & \multicolumn{1}{c}{\rotatebox{90}{Fe5782}} & \multicolumn{1}{c}{\rotatebox{90}{Na5895}} & \multicolumn{1}{c}{\rotatebox{90}{TiO$_1$}} & \multicolumn{1}{c}{\rotatebox{90}{TiO$_2$}}\\ \hline
$\tilde{\rm D}_{4000}$ & W94 & \ldots & $0.2$ & $0.2$ & $1.7$ & $0.5$ & $0.4$ & $0.7$ & $0.8$ & $0.0$ & $0.3$ & $0.4$ & $0.3$ & $0.2$ & $1.0$ & $0.6$ & $0.4$ & $0.5$ & $0.5$ & $0.6$ & $0.4$ & $1.0$ & $0.6$\\ 
$\tilde{\rm D}_{4000}$ & V00 & \ldots & $0.3$ & $0.4$ & $0.9$ & $1.3$ & $0.4$ & $0.7$ & $1.1$ & $0.0$ & $0.4$ & $0.5$ & $0.7$ & $0.4$ & $0.7$ & $0.8$ & $0.6$ & $0.7$ & $0.6$ & $0.8$ & $0.6$ & $0.8$ & $1.1$\\ 
$\tilde{\rm D}_{4000}$ & BC01 & \ldots & $0.2$ & $0.4$ & $1.2$ & $1.9$ & $0.3$ & $0.6$ & $0.8$ & $0.0$ & $0.5$ & $0.4$ & $0.2$ & $0.0$ & $0.5$ & $0.6$ & $0.4$ & $0.5$ & $0.6$ & $0.7$ & $0.3$ & $1.4$ & $0.9$\\ 
\hline
Fe4668 & W94 & $0.0$ & $1.1$ & $1.2$ & $1.2$ & $1.1$ & $1.3$ & $1.6$ & $1.7$ & \ldots & $1.2$ & $2.2$ & $1.0$ & $0.8$ & $1.2$ & $1.7$ & $1.9$ & $1.9$ & $2.4$ & $3.0$ & $1.3$ & $1.6$ & $1.4$\\ 
Fe4668 & V00 & $0.0$ & $0.9$ & $0.9$ & $1.1$ & $1.3$ & $1.2$ & $1.7$ & $1.5$ & \ldots & $1.2$ & $2.1$ & $0.8$ & $0.7$ & $1.3$ & $1.5$ & $1.7$ & $1.7$ & $3.1$ & $2.2$ & $1.1$ & $1.6$ & $1.3$\\ 
Fe4668 & BC01 & $0.0$ & $0.9$ & $0.9$ & $1.2$ & $1.3$ & $1.4$ & $1.7$ & $1.6$ & \ldots & $1.2$ & $2.4$ & $0.9$ & $0.8$ & $1.3$ & $1.6$ & $1.9$ & $1.8$ & $2.3$ & $3.1$ & $1.2$ & $1.5$ & $1.2$\\ 
\hline
Mg$_2$ & W94 & $0.2$ & $1.5$ & $1.4$ & $1.4$ & $1.0$ & $1.5$ & $1.9$ & $2.2$ & $0.8$ & $0.9$ & $1.3$ & $2.0$ & \ldots & $1.6$ & $1.6$ & $1.3$ & $1.4$ & $1.3$ & $1.4$ & $1.5$ & $1.5$ & $2.0$\\ 
Mg$_2$ & V00 & $0.4$ & $1.2$ & $1.4$ & $1.2$ & $1.5$ & $1.3$ & $1.5$ & $2.4$ & $0.7$ & $0.9$ & $1.2$ & $1.7$ & \ldots & $1.8$ & $1.7$ & $1.3$ & $1.5$ & $1.3$ & $1.5$ & $1.7$ & $1.4$ & $1.4$\\ 
Mg$_2$ & BC01 & $0.0$ & $1.9$ & $1.4$ & $1.3$ & $1.3$ & $1.4$ & $1.7$ & $2.5$ & $0.8$ & $1.0$ & $1.3$ & $1.8$ & \ldots & $2.8$ & $1.8$ & $1.3$ & $1.5$ & $1.3$ & $1.5$ & $1.6$ & $1.5$ & $1.4$\\ 
\hline
Na5895 & W94 & $0.4$ & $2.1$ & $2.6$ & $1.6$ & $1.3$ & $3.1$ & $2.9$ & $2.6$ & $1.3$ & $1.3$ & $1.8$ & $1.7$ & $1.5$ & $1.7$ & $2.4$ & $2.0$ & $2.1$ & $1.8$ & $1.9$ & \ldots & $1.8$ & $2.2$\\ 
Na5895 & V00 & $0.6$ & $2.0$ & $2.6$ & $1.5$ & $1.8$ & $1.9$ & $2.0$ & $2.8$ & $1.1$ & $1.3$ & $1.7$ & $1.6$ & $1.7$ & $3.7$ & $2.5$ & $1.8$ & $2.0$ & $1.8$ & $1.9$ & \ldots & $1.7$ & $1.7$\\ 
Na5895 & BC01 & $0.3$ & $1.6$ & $1.5$ & $1.5$ & $1.6$ & $2.1$ & $2.5$ & $2.6$ & $1.2$ & $1.3$ & $1.7$ & $1.6$ & $1.6$ & $2.0$ & $2.7$ & $1.8$ & $2.0$ & $1.7$ & $1.9$ & \ldots & $1.8$ & $1.6$\\ 
\hline
\end{tabular}